\documentclass[nofootinbib,reprint,groupedaddress,preprintnumbers]{revtex4-2}

\usepackage[bottom]{footmisc}
\usepackage{amsmath,amssymb,marginnote}
\usepackage{multirow} 
\usepackage{booktabs} 
\usepackage{graphicx,subfigure}
\usepackage{dcolumn}
\usepackage{bm}
\usepackage[mathlines]{lineno}
\usepackage[colorlinks=true, allcolors=blue]{hyperref}
\usepackage[usenames]{color}
\usepackage{xcolor}
\usepackage{comment}
\usepackage{mathrsfs}

\begin{document}

\preprint{LA-UR-25-25004}

\title{Local-equilibrium theory of neutrino oscillations}

\author{Lucas Johns}
\email{ljohns@lanl.gov}
\affiliation{Theoretical Division, Los Alamos National Laboratory, Los Alamos, NM 87545, USA}

\author{Anson Kost}
\affiliation{Department of Physics and Astronomy, University of New Mexico, Albuquerque, NM 87131, USA}

\begin{abstract}
Miscidynamics describes coarse-grained neutrino transport under the assumption that flavor mixing is in local equilibrium. Here we introduce the concept of turbulent flavor-wave viscosity and develop techniques for including it in miscidynamics. This extension of the theory is necessary when neutrinos develop weak flavor instabilities as a result of astrophysical driving. The flavor-wave dispersion relation is obtained using a new and more general form of linear analysis, which does not require small flavor coherence. Flavor-wave transport is then approximated using wave kinetics and geometric optics. We hypothesize that the dynamical emergence of local oscillation invariants restricts the extent of flavor thermalization. In this view, flavor evolution is sculpted by both entropy-increasing and order-promoting factors.
\end{abstract}

\maketitle

\section{Introduction}

Theoretical evidence increasingly suggests that neutrino oscillations are important for the dynamical and chemical outcomes of core-collapse supernovae (CCSNe) and neutron star mergers (NSMs) \cite{sawyer2016neutrino, wu2015effects, wu2017fast, 
wu2017imprints, abbar2019occurrence, delfan2019linear, morinaga2020fast, xiong2020potential, ko2020neutrino, george2020fast, abbar2021characteristics, nagakura2021where, tamborra2021new, johns2021collisional, li2021, just2022, fernandez2022, fujimoto2023explosive, xiong2023collisional, liu2023universality, ehring2023, ehring2023fast, ehring2024gravitational, froustey2024neutrino, akaho2024collisional, liu2024muon, mukhopadhyay2024time, qiu2025neutrino, wang2025, lund2025angle, mori2025, abbar2025using}. The accurate incorporation of flavor mixing into astrophysical simulations has consequently become a pressing issue at the forefront of particle astrophysics \cite{volpe2024, tamborra2024, johns2025neutrino}.

The central challenge is that the quantum kinetic equation (QKE) \cite{sigl1993general, vlasenko2014neutrino, stirner2018liouville, richers2019neutrino}
\begin{equation}
    i ( \partial_t + \bm{\hat{q}} \cdot \partial_{\bm{r}} ) \rho_{\bm{q}} = [H_{\bm{q}}, \rho_{\bm{q}}] + i C_{\bm{q}} \label{eq:qke}
\end{equation}
is computationally intractable in these extreme stellar environments. Here $\rho_{\bm{q}}$ is the density matrix of neutrinos at momentum $\bm{q}$, $H_{\bm{q}}$ is the Hamiltonian, and $iC_{\bm{q}}$ is the collision term. A fervent effort is now underway to develop the theoretical ideas and methods required to approximate neutrino transport in a way that is accurate yet reasonably inexpensive \cite{johns2020neutrino, johns2020fast, bhattacharyya2020late, bhattacharyya2020fast, wu2021collective, richers2021neutrino, nagakura2022general, xiong2022evolution, shalgar2022neutrino, shalgar2023neutrino, nagakura2023bhatnagar, johns2023thermodynamics, johns2024subgrid, kost2025once, xiong2024fast, xiong2025robust, fiorillo2024collisions, fiorillo2024fast,
fiorillo2025collective, liu2024quasi, 
abbar2024physics, richers2024asymptotic, kneller2025quantum, froustey2025quantum, zaizen2025spectral, nagakura2025resolution, liu2025asymptotic}.

The focus of this work is miscidynamics, a coarse-grained theory of neutrino transport that self-consistently leverages the separation between oscillation and astrophysical scales \cite{johns2023thermodynamics, johns2024subgrid, kost2025once}. Roughly speaking, the oscillation length scale is $\Sigma^{-1}$, where $\Sigma \sim G_F n$ is the energy scale of forward scattering in the medium and $n$ is a representative particle number density. In some regions of interest, $\Sigma^{-1}$ translates to a sub-millimeter length scale---far below the kilometer scale typical of CCSNe and NSMs. The aim of miscidynamics is to draw a distinction between macroscopic (grid-level) and mesoscopic (subgrid) physics and apply approximations accordingly.

The approximation that defines miscidynamics is the assumption of local mixing equilibrium. This condition means that the coarse-grained density matrix $\langle \rho_{\bm{q}} \rangle$ is close to a state of mixing equilibrium $\rho_{\bm{q}}^{\textrm{eq}}$ defined by
\begin{equation}
    [ H_{\bm{q}}^{\textrm{eq}}, \rho_{\bm{q}}^{\textrm{eq}} ] = 0. \label{eq:mixeq0}
\end{equation}
The superscript on $H^{\textrm{eq}}_{\bm{q}}$ indicates that $H_{\bm{q}}$, which depends on $\rho_{\bm{q}'}$ for all $\bm{q}'$ through neutrino--neutrino forward scattering, is evaluated using $\rho^{\textrm{eq}}_{\bm{q}'}$. If $\langle \rho_{\bm{q}} \rangle = \rho_{\bm{q}}^{\textrm{eq}}$ exactly, then the uncorrelated part of the coarse-grained commutator vanishes: $[\langle H_{\bm{q}} \rangle, \langle \rho_{\bm{q}} \rangle] = 0$. If the correlated part vanishes as well, either because subgrid fluctuations are negligible or effectively random, then oscillations are fully absorbed into the local-equilibrium condition. These two assumptions---local mixing equilibrium and the vanishing of subgrid correlations---ensure a successful application of coarse-graining to Eq.~\eqref{eq:qke}. 

The next step in the program is to allow some amount of leeway in these assumptions. Contributions from nonequilibrium deviations,
\begin{equation}
    [\langle H_{\bm{q}} \rangle, \langle \rho_{\bm{q}} \rangle] \neq 0,
\end{equation}
and nonzero correlations,
\begin{equation}
    \langle [H_{\bm{q}}, \rho_{\bm{q}}] \rangle - [\langle H_{\bm{q}} \rangle, \langle \rho_{\bm{q}} \rangle] \neq 0, \label{eq:nonzerocorr}
\end{equation}
are both tolerable as long as they only matter for the grid-level dynamics and can still be neglected on subgrid scales. In this work we formulate the grid-level corrections arising from subgrid correlations, deferring nonequilibrium deviations to future work.

To understand the philosophy of miscidynamics, it is helpful to draw an analogy with hydrodynamics \cite{johns2024subgrid}. For a gas in local thermodynamic equilibrium, the local distribution function is approximately a Maxwellian $f_{\bm{q}}^\textrm{MB}$ (for example) and the coarse-grained evolution is hydrodynamic. For neutrinos in local mixing equilibrium, the local distribution function is approximately $\rho_{\bm{q}}^{\textrm{eq}}$ and the coarse-grained evolution is miscidynamic. In the same way that the Boltzmann collision term vanishes at $f_{\bm{q}} = f_{\bm{q}}^\textrm{MB}$, the oscillation term $[H_{\bm{q}}, \rho_{\bm{q}}]$ vanishes at $\rho_{\bm{q}} = \rho_{\bm{q}}^{\textrm{eq}}$. Both theories, hydrodynamics and miscidynamics, describe the transport of microscopically conserved quantities over macroscopic scales.

In miscidynamics, local mixing equilibrium is determined by the local values of the \textit{oscillation invariants}, the coarse-grained quantities that are unchanged by oscillations. Restricting our attention to two-flavor mixing, as we do throughout the paper, we expand $\rho_{\bm{q}}$ using Pauli matrices:
\begin{equation}
    \rho_{\bm{q}} = \frac{P^0_{\bm{q}}+\bm{P}_{\bm{q}}\cdot\bm{\sigma}}{2}. \label{eq:pauli}
\end{equation}
Mixing equilibrium is most notably a function of the coarse-grained flavor polarizations $|\langle \bm{P}_{\bm{q}} \rangle|$. These invariants are dynamically emergent and not strictly conserved under oscillations.

In previous work, miscidynamics was explicitly presented only in the adiabatic limit. The adiabatic approximation imposes the assumption of vanishing subgrid correlations and thus forbids any change in $|\langle \bm{P}_{\bm{q}} \rangle|$ stemming from the oscillation term $[ H_{\bm{q}}, \rho_{\bm{q}}]$. Adiabaticity is plausible, though emphatically not guaranteed, if the separation between the oscillation scale $l_{\textrm{osc}}$ and the astrophysical (coarse-graining) scale $l_{\textrm{astro}}$ is infinite. (We will be using length and time scales interchangeably.) Scale separation is quantified by the dimensionless parameter \cite{johns2024subgrid}
\begin{equation}
    \varepsilon \equiv \frac{l_{\textrm{osc}}}{l_{\textrm{astro}}},
\end{equation}
the analogue of the Knudsen number in hydrodynamics. The theory analogous to adiabatic ($\varepsilon = 0$) miscidynamics is inviscid hydrodynamics as governed by the Euler equations. In the latter theory, there is no vestige of particle collisions at the hydrodynamic scale apart from the maintenance of local thermodynamic equilibrium. In adiabatic miscidynamics, oscillations are fully subsumed by the condition of local mixing equilibrium.

In reality, $\varepsilon$ is an imprecise quantity because there is no single scale characterizing either the oscillation dynamics or the astrophysics. At the extreme, some features of the oscillation dynamics may flout scale separation altogether. Subgrid correlations [Eq.~\eqref{eq:nonzerocorr}] can nonadiabatically alter $|\langle \bm{P}_{\bm{q}} \rangle|$ even in the limit of infinite scale separation. If the strength of correlations is set by the astrophysics (\textit{i.e.}, the neutrino opacities and emissivities and the temporal and spatial variation of the fluid), then their influence, despite originating from $[H_{\bm{q}},\rho_{\bm{q}}]$, is hitched to $l_{\textrm{astro}}$. Their effect is nonvanishing and must be included as an additional grid-level term even if we take
\begin{equation}
    \varepsilon \sim \frac{\Sigma^{-1}}{l_{\textrm{astro}}} \rightarrow 0.
\end{equation}
Here we are concerned with precisely this phenomenon: nonadiabaticity in the \textit{quasistatic} ($\varepsilon \rightarrow 0$) limit. As we have said, we do not undertake a detailed treatment of nonadiabaticity due to finite rates of change and nonequilibrium deviations ($\varepsilon \neq 0$).

Suppose that $\langle \rho_{\bm{q}} \rangle$ is the spatial average over a cubic region with side length $l_{\textrm{box}} \lesssim l_{\textrm{astro}}$. Then inside the region we have
\begin{equation}
    \rho_{\bm{q}} = \langle \rho_{\bm{q}} \rangle + \sum_{\bm{n} \neq \bm{0}} \rho_{\bm{q},\bm{n}} e^{i \bm{k}_{\bm{n}}\cdot\bm{r}}
\end{equation}
with $\bm{k}_{\bm{n}} = 2 \pi \bm{n}/l_{\textrm{box}}$, $\bm{n}$ being a vector of integers. Because $\rho_{\bm{q}}(t,\bm{r})$ is Hermitian, $\rho_{\bm{q},-\bm{n}} = \rho_{\bm{q},\bm{n}}^\dagger$. We use the term \textit{flavor waves} to refer to the subgrid ($\bm{n} \neq \bm{0}$) modes of the neutrino radiation field. In the theory presented here, nonadiabaticity is attributed to \textit{turbulent flavor-wave viscosity}, in analogy with the turbulent eddy viscosity in a fluid. The mean flavor distributions $\langle \rho_{\bm{q}} \rangle$ are constantly being driven by collisional processes and neutrino advection. Flavor waves respond to this driving. Under certain conditions, such as when $\langle \rho_{\bm{q}} \rangle$ is driven through weakly unstable or marginally stable mixing equilibria \cite{fiorillo2024fast, liu2024quasi, xiong2025robust}, flavor waves exert a kind of ``friction'' resisting the grid-level changes. This is how an $l_{\textrm{astro}}$ scale emerges in the oscillation dynamics. The viscosity is, in a loose sense, proportional to the driving.

As we show below, the miscidynamic equation for $\bm{P}_{\bm{q}}$ in the quasistatic limit is
\begin{equation}
    ( \partial_t + \bm{\hat{q}} \cdot \partial_{\bm{r}} ) \bm{P}^{\textrm{eq}}_{\bm{q}} = -(\gamma_{\bm{q}}  + \Gamma_{\bm{q}})\bm{P}^{\textrm{eq}}_{\bm{q}},
\end{equation}
where $\gamma_{\bm{q}}$ is the turbulent flavor-wave viscosity and $\Gamma_{\bm{q}}$ is the collisional depolarization rate. (This equation is written in a specific $t$-, $\bm{r}$-, and $\bm{q}$-dependent basis that we call the \textit{local equilibrium frame}.) In an exact description, $\gamma_{\bm{q}}$ depends on the detailed evolution of every $ \bm{P}_{\bm{q},\bm{n}}$. But obtaining this would be nearly as expensive as solving the QKE in its entirety. A closure is needed to maintain the coarse-grained quality of miscidynamics. The closure problem encountered here mirrors the one that arises in modeling the Reynolds stress of a turbulent fluid.

To approximate $\gamma_{\bm{q}}$, we develop three main innovations, which we summarize below.

\textbf{\textit{Weakly coupled flavor waves.}} A great deal of work in this research area is based on the stipulation that $\rho_{\bm{q}}$ is nearly diagonal in the flavor basis at all $\bm{r}$. This assumption allows for linearization of the QKE in the small off-diagonal elements \cite{banerjee2011, izaguirre2017}. We introduce a more general linearization procedure based on the assumption of \textit{weak inhomogeneity}: flavor-wave polarizations $\bm{P}_{\bm{q},\bm{n}\neq\bm{0}}$ are assumed to be small relative to the slowly varying background polarizations $\langle\bm{P}_{\bm{q}}\rangle$. This more general starting point allows us to formulate flavor waves as $\bm{n} \neq \bm{0}$ excitations on any mixing-equilibrium background, not just the particular equilibrium that has all $\langle\bm{P}_{\bm{q}}\rangle$ (anti)aligned with the flavor axis $\bm{z}$.

Flavor waves are emergent quasiparticles, collective neutrino excitations analogous to phonons, magnons, and plasmons. (For other work inspired by these parallels, see, \textit{e.g.}, Refs.~\cite{duan2009symmetries, duan2021flavor, fiorillo2024theory, fiorillo2025collective}.) Flavor waves are weakly coupled to each other under conditions of weak inhomogeneity. Linearization allows us to determine the dispersion relation $\Omega_{\bm{n},i}$ relating the ``energy'' $\Omega^{\textrm{R}} \equiv \textrm{Re}(\Omega)$ and decay/growth rate $\Omega^{\textrm{I}} \equiv \textrm{Im}(\Omega)$ of a flavor wave to its ``momentum'' $\bm{k}_{\bm{n}}$. The $i$ subscript above identifies a particular band (in the sense of band structure in condensed matter). There are three times as many bands as there are momenta $\bm{q}$, with the factor of three coming from the number of polarization vector components. The $t_{\textrm{osc}}$-scale flavor-wave dynamics follows from an eigendecomposition.

\textbf{\textit{Flavor-wave kinetics.}} We formulate a theory of flavor-wave kinetics to approximate the $t_{\textrm{astro}}$-scale evolution. In this we are following the vision laid out previously of solving the flavor-wave closure problem in a manner resembling the closure of the Bogoliubov--Born--Green--Kirkwood--Yvon (BBGKY) hierarchy by kinetic theory \cite{johns2023thermodynamics, johns2024subgrid}.

Flavor-wave kinetics follows from application of the \textit{secular approximation}. Phase information is only important inasmuch as it influences the grid-level dynamics through the nonlinear terms. (Throughout this paper, \textit{nonlinear} refers to terms that are quadratic in $\rho_{\bm{q},\bm{n\neq\bm{0}}}$ or $\bm{P}_{\bm{q},\bm{n}\neq\bm{0}}$.) For example, phases appear in integrals over $t$ with integrands containing phase factors of the form
\begin{equation}
    e^{i (\Omega^{\textrm{R}}_{\bm{n}, i} -\Omega^{\textrm{R}}_{\bm{n}-\bm{m}, j} - \Omega^{\textrm{R}}_{\bm{m}, l})t},
\end{equation}
with the integral taken over a $t_{\textrm{astro}}$ interval. Under the secular approximation, these integrals vanish unless a resonance condition is met:
\begin{equation}
    \Omega^{\textrm{R}}_{\bm{n}, i} -\Omega^{\textrm{R}}_{\bm{n}-\bm{m}, j} - \Omega^{\textrm{R}}_{\bm{m}, l} = 0.
\end{equation}
All other integrals average out to zero due to the scale separation $t_{\textrm{astro}} \gg \Sigma^{-1}$ because $\Sigma$ is the typical scale of $\Omega^{\textrm{R}}$.

In flavor-wave kinetics, $\gamma_{\bm{q}}$ is approximated as the net decay rate of neutrinos (\textit{i.e.}, $\langle\bm{P}_{\bm{q}}\rangle$) into flavor waves---or, if $\gamma_{\bm{q}} < 0$, the net fusion rate of flavor waves into neutrinos. Flavor waves additionally participate in three-wave scattering processes, which reshape and potentially thermalize the distribution of flavor waves over $\bm{k}$. Both types of processes, (inverse) neutrino decay and three-wave scattering, obey energy and momentum conservation.

\textbf{\textit{Flavor-wave propagation.}} Kinetics approximates the local dynamics of flavor waves. We also need some way of describing their transport over long durations and macroscopic distances. There are two aspects to this. First, we need a prescription for moving flavor waves between the coarse-graining cells in which they are defined. We use the \textit{ray approximation}, according to which flavor waves propagate with group velocities and forces determined from Hamilton's equations in $(\bm{r},\bm{k})$ phase space. Second, the flavor-wave eigensystems change as the coarse-grained polarizations $\langle\bm{P}_{\bm{q}}\rangle$ slowly vary in $t$ and $\bm{r}$. We present the \textit{quasistatic approximation} as the simplest way of handling this variation, though we caution that non-quasistatic transitions between bands $\Omega_{\bm{n},i}$ and $\Omega_{\bm{n},j}$ may need to be accounted for. We leave aside the problem of formulating simple approximations for such variation-driven transitions, which are probably enhanced by the occurrence of exceptional points in the flavor-wave eigensystems.

Through the sequence of approximations sketched above, we arrive at an extension of miscidynamics that transports flavor-wave distributions in addition to mixing equilibria. It accounts for turbulent flavor-wave viscosity without resolving the fine-grained evolution of the flavor waves.

Many of the ideas and techniques we develop are not particular to miscidynamics, even though we develop them within this framework. For example, all of the following can be employed without assuming local mixing equilibrium: 
\begin{itemize}
    \item weak-inhomogeneity linear analysis (generalizing small-coherence linear analysis),
    \item flavor polarization waves (generalizing flavor coherence waves),
    \item flavor-wave kinetics (derived here by applying the secular approximation to isolate resonant interactions),
    \item the quasistatic approximation for flavor-wave evolution (as a simplest possible treatment of how a $t$- and $\bm{r}$-varying background affects the distributions and properties of flavor waves), and
    \item the ray approximation for flavor-wave propagation (for bridging the gap between calculations in local periodic boxes and global astrophysical environments).
\end{itemize}
These are all approximations aimed at the description of flavor-wave dynamics. What defines miscidynamics is how it approximates the evolution of spatially averaged ($\bm{n} = \bm{0}$) quantities. If $\bm{\hat{P}}_{\bm{q},\bm{0}} = \pm \bm{z}$ everywhere and for all $\bm{q}$, and if we assume either that $\bm{B} = \pm \bm{z}$ or $\omega = 0$ (mass basis vector $\bm{B}$ and vacuum oscillation frequency $\omega$), then the condition of local mixing equilibrium is trivially satisfied. But miscidynamics is concerned with the possibility that these are not always good approximations in practice. They are obviously not applicable to MSW resonant conversion \cite{mikheyev1985resonance} or spectral swaps \cite{duan2006simulation, duan2006coherent, raffelt2007adiabaticity, raffelt2007self2}, for instance. The objective is to develop a theory encompassing all forms of flavor mixing that are realized in astrophysics.

Toward the end of the paper, we expand on another theory that has interconnections with miscidynamics:

\textbf{\textit{Neutrino quantum thermodynamics.}} The original presentation of miscidynamics used thermodynamic arguments as motivation for the hypothesis of local mixing equilibrium \cite{johns2023thermodynamics}. These are in fact two separate theories, and miscidynamics can be formulated without any recourse to thermodynamics or related concepts such as ergodicity \cite{johns2023collisional, johns2024ergodicity}. Nonetheless, thermodynamics sheds light on neutrino flavor evolution, and there are important open questions regarding their relationship. For example, to what extent does coarse-grained flavor evolution adhere to some kind of $H$-theorem? To what degree is flavor ergodicity a useful approximation? Is neutrino--neutrino forward scattering capable of thermalizing neutrino flavor polarizations and/or flavor-wave populations?

We revisit the \textit{flavor-temperature hypothesis} of Ref.~\cite{johns2023thermodynamics} and propose the alternative that flavor thermalization is generally obstructed by the dynamical emergence of approximate invariants. We also discuss the concept of \textit{flavor-wave equilibrium}, which is distinct from mixing equilibrium, and speculate that the post-instability high-$|\bm{k}|$ exponential scaling of the polarization power spectrum \cite{bhattacharyya2020late, bhattacharyya2020fast, wu2021collective, richers2021neutrino, richers2022code, urquilla2024chaos, padilla2025flavor} may reflect equilibration to the maximum-entropy flavor-wave distribution. Fleshing out this idea will require the extension of neutrino quantum thermodynamics to encompass flavor waves \cite{fiorillo2025dispersion}.

In Sec.~\ref{sec:mixeq} we define and discuss various aspects of mixing equilibrium. In Sec.~\ref{sec:misc} we present miscidynamics and the closure problem associated with turbulent flavor-wave viscosity. In Sec.~\ref{sec:neareq} we introduce the weak-inhomogeneity approximation and the formalism of weakly coupled flavor waves. In Sec.~\ref{sec:fwdynamics} we develop a theory of flavor-wave dynamics. In Sec.~\ref{sec:thermo} we adopt the thermodynamic perspective. In Sec.~\ref{sec:relation} we relate the flavor-wave transport theory developed here to the quasilinear theory put forward by Fiorillo and Raffelt \cite{fiorillo2024fast, fiorillo2025collective, fiorillo2026flavomon, fiorillo2026quasi}. In Sec.~\ref{sec:guide} we summarize our main proposals in a way geared toward numerical implementation, skipping over the derivations and nuances covered in the rest of the paper. Finally, in Sec.~\ref{sec:summ} we summarize our work.

\section{Mixing equilibrium\label{sec:mixeq}}

\subsection{Definition\label{sec:definition}}

The building blocks of miscidynamics are mixing equilibria \cite{johns2023thermodynamics, johns2024subgrid}, states for which the coarse-grained density matrices $\langle \rho_{\bm{q}} \rangle$ at all $\bm{q}$ are simultaneously static under their coarse-grained Hamiltonians $\langle H_{\bm{q}} \rangle$ [Eq.~\eqref{eq:mixeq0}]. When this condition is met, the polarization vectors satisfy
\begin{equation}
    \langle \bm{H}_{\bm{q}} \rangle \times \langle \bm{P}_{\bm{q}} \rangle = 0. \label{eq:mixeqHP}
\end{equation}
In mixing equilibrium, every $\langle \bm{P}_{\bm{q}} \rangle$ is (anti)aligned with the corresponding $\langle \bm{H}_{\bm{q}} \rangle$. This does not imply that all $\langle \bm{P}_{\bm{q}} \rangle$ are (anti)aligned with one another. Mixing equilibrium can also be expressed as
\begin{equation}
    \epsilon_{\bm{q}} \equiv \frac{\langle \bm{P}_{\bm{q}} \rangle \cdot \langle \bm{H}_{\bm{q}} \rangle}{|\langle \bm{P}_{\bm{q}} \rangle | | \langle \bm{H}_{\bm{q}} \rangle|} = \pm 1. \label{eq:epsmixeq}
\end{equation}
These various ways of defining mixing equilibrium should be understood as self-consistency conditions on the coarse-grained neutrino flavor distributions, because $\bm{H}_{\bm{q}}$ at some particular momentum $\bm{q}$ depends on $\bm{P}_{\bm{q}'}$ at all other momenta $\bm{q}'$.

The collectivity is due to the neutrino--neutrino forward-scattering term in the Hamiltonian:
\begin{equation}
    \bm{H}^{\nu\nu}_{\bm{q}} = \sqrt{2} G_F ( \bm{D}_0 - \bm{\hat{q}}\cdot \bm{\vec{D}}_1 ), \label{eq:Hnunu}
\end{equation}
which we have written in terms of the collective polarization vectors
\begin{align}
    &\bm{D}_0 \equiv \int \frac{d^3 \bm{q}'}{(2\pi)^3} ( \bm{P}_{\bm{q}'} - \bm{\bar{P}}_{\bm{q}'}), \notag \\
    &\bm{\vec{D}}_1 \equiv \int \frac{d^3 \bm{q}'}{(2\pi)^3} \bm{\hat{q}}' ( \bm{P}_{\bm{q}'} - \bm{\bar{P}}_{\bm{q}'}).
\end{align}
Overbars denote antineutrino quantities, and the vector arrow on $\bm{\vec{D}}_1$ emphasizes the fact that it has a vector structure in both coordinate space and flavor space.

The self-consistency conditions can also be cast in the form
\begin{widetext}
\begin{equation}
    \langle \bm{D}_0 \rangle = \int \frac{d^3 \bm{q}'}{(2\pi)^3} \bigg( \frac{\epsilon_{\bm{q}'} |\langle\bm{P}_{\bm{q}'}\rangle| \langle\bm{H}_{\bm{q}'}\rangle}{|\langle\bm{H}_{\bm{q}'}\rangle|} - \frac{\bar{\epsilon}_{\bm{q}'} |\langle\bm{\bar{P}}_{\bm{q}'}\rangle| \langle\bm{\bar{H}}_{\bm{q}'}\rangle}{|\langle\bm{\bar{H}}_{\bm{q}'}\rangle|} \bigg), ~~~ \langle\bm{\vec{D}}_1\rangle = \int \frac{d^3 \bm{q}'}{(2\pi)^3} \bm{\hat{q}}' \bigg( \frac{\epsilon_{\bm{q}'} |\langle\bm{P}_{\bm{q}'}\rangle| \langle\bm{H}_{\bm{q}'}\rangle}{|\langle\bm{H}_{\bm{q}'}\rangle|} - \frac{\bar{\epsilon}_{\bm{q}'} |\langle\bm{\bar{P}}_{\bm{q}'}\rangle| \langle\bm{\bar{H}}_{\bm{q}'}\rangle}{|\langle\bm{\bar{H}}_{\bm{q}'}\rangle|}\bigg). \label{eq:selfcons}
\end{equation}
\end{widetext}
These follow from requiring the collective parts of the Hamiltonian to be consistent with $\bm{P}_{\bm{q}}$ satisfying Eq.~\eqref{eq:epsmixeq}. The quantities on the left-hand sides of Eqs.~\eqref{eq:selfcons} implicitly appear on the right-hand sides through $\langle \bm{H}_{\bm{q}'}\rangle$ and $\langle \bm{\bar{H}}_{\bm{q}'}\rangle$, via Eq.~\eqref{eq:Hnunu}.

In miscidynamics, as we explain below, one assumes knowledge of $|\langle \bm{P}_{\bm{q}} \rangle|$ and $\epsilon_{\bm{q}}$ in determining the local mixing-equilibrium state. Equilibrium is thus found by holding fixed the magnitudes of the vectors $\langle \bm{P}_{\bm{q}} \rangle$ and varying their orientations. Reorienting $\langle \bm{P}_{\bm{q}}\rangle$ changes $\langle \bm{H}_{\bm{q}'}\rangle$. While Eqs.~\eqref{eq:selfcons} are written as self-consistency conditions for $\langle \bm{D}_0\rangle$ and $\langle\bm{\vec{D}}_1\rangle$, a procedural way of thinking about these equations is that each $\langle \bm{P}_{\bm{q}}\rangle$ must be rotated in flavor space until it is appropriately (anti)aligned with $\langle \bm{H}_{\bm{q}}\rangle$, as prescribed by $\epsilon_{\bm{q}}$.

We use the symbols $\rho^{\textrm{eq}}_{\bm{q}}$ and $\bm{P}^{\textrm{eq}}_{\bm{q}}$ to denote $\langle \rho_{\bm{q}} \rangle$ and $\langle \bm{P}_{\bm{q}} \rangle$ satisfying the condition of mixing equilibrium.

\subsection{Dynamical emergence of coarse-grained oscillation invariants}

The particular state of mixing equilibrium that obtains in some region is determined by the external parameters that influence $\bm{H}_{\bm{q}}$ (\textit{e.g.}, electron density $n_e$) and by a set of parameters that we call the \textit{oscillation invariants}. The oscillation invariants are quantities that pertain to the neutrinos themselves and are constant under flavor mixing, with caveats that we explain in this and the next subsection.

The oscillation invariants first of all include the traces $\langle P^0_{\bm{q}} \rangle$ because they are exactly conserved under the collisionless QKE in an isolated box. Throughout this paper we assume that subgrid inhomogeneity arises only from $\bm{P}_{\bm{q}}$, so that $P^0_{\bm{q}} = \langle P^0_{\bm{q}} \rangle$ at all locations within the region. Miscidynamics could be extended to allow for subgrid inhomogeneity in the densities of neutrinos and other particles, but we do not pursue that here.

The other oscillation invariants in miscidynamics are $|\langle \bm{P}_{\bm{q}} \rangle|$ and $\epsilon_{\bm{q}}$. This requires explanation because it is not immediately obvious that they should be treated as invariants. Consider the evolution of $\langle \bm{P}_{\bm{q}} \rangle$ in an isolated region:
\begin{equation}
    \frac{d\langle\bm{P}_{\bm{q}}\rangle}{dt} = \langle \bm{H}_{\bm{q}} \times \bm{P}_{\bm{q}}\rangle + \langle \bm{C}_{\bm{q}}\rangle,
\end{equation}
where $\bm{C}_{\bm{q}}$ is the flavor-space vector defined from the collision term $iC_{\bm{q}}$ in Eq.~\eqref{eq:qke}. In identifying oscillation invariants, we focus on the short-time evolution,  ignoring collisions and any other astrophysical driving. Hence we examine
\begin{equation}
    \frac{d|\langle\bm{P}_{\bm{q}}\rangle|}{dt} = \frac{\langle\bm{P}_{\bm{q}}\rangle \cdot \langle\bm{H}_{\bm{q}}\times\bm{P}_{\bm{q}} \rangle}{|\langle\bm{P}_{\bm{q}}\rangle|}.
\end{equation}
The right-hand side does not vanish in general, and so $|\langle\bm{P}_{\bm{q}}\rangle|$ is not strictly invariant under oscillations.

But consider separating $\langle\bm{H}_{\bm{q}}\times\bm{P}_{\bm{q}} \rangle$ into a factorized part and a remainder $\bm{c}_{\bm{q}}$:
\begin{equation}
    \bm{c}_{\bm{q}} \equiv \langle\bm{H}_{\bm{q}}\times\bm{P}_{\bm{q}} \rangle - \langle\bm{H}_{\bm{q}}\rangle\times\langle\bm{P}_{\bm{q}} \rangle.
\end{equation}
In the case that $\bm{c}_{\bm{q}} = 0$, we have
\begin{equation}
    |\langle\bm{P}_{\bm{q}}\rangle| = \textrm{constant}.
\end{equation}
Factorization results in the dynamical emergence of $|\langle\bm{P}_{\bm{q}}\rangle|$ as a coarse-grained oscillation invariant. We use the symbol $\bm{c}_{\bm{q}}$ because this quantity represents subgrid \textit{correlations}. As we will see later in the context of flavor-wave kinetics, $\bm{c}_{\bm{q}}$ also represents the grid-level effects of flavor-wave \textit{collisions} in the form of neutrino decay to flavor waves and the inverse reaction. Similarly, $iC_{\bm{q}}$ represents particle collisions as well as the effects of many-body correlations \cite{froustey2020neutrino, johns2023neutrino, kost2024once}. The emergence of $\epsilon_{\bm{q}}$ as an invariant, on the other hand, hinges on a generalization of adiabatic tracking at the global, coarse-grained level.

To be more specific about the nature of subgrid correlations, we decompose $\bm{c}_{\bm{q}}$ into contributions from individual Fourier modes:
\begin{equation}
    \bm{c}_{\bm{q}} = \sum_{\bm{n}\neq\bm{0}} \bm{c}_{\bm{q},\bm{n}} = \sum_{\bm{n}\neq\bm{0}} \bm{H}_{\bm{q},-\bm{n}} \times \bm{P}_{\bm{q},\bm{n}}. \label{eq:cqdefn}
\end{equation}
Here we see that subgrid correlations reflect the distribution of flavor waves. Obviously one way to ensure $\bm{c}_{\bm{q}} = 0$ is to have vanishing $\bm{P}_{\bm{q},\bm{n}}$ at all $\bm{n}$. Less trivially, correlations vanish if
\begin{equation}
    \bm{H}_{\bm{q},-\bm{n}} \times \bm{P}_{\bm{q},\bm{n}} = 0
\end{equation}
for all $\bm{n}$. This condition extends the concept of mixing equilibrium to $\bm{n} \neq \bm{0}$, though we will continue to use the term to refer to $\bm{n} = \bm{0}$ only. The fact that nonvanishing $\bm{P}_{\bm{q},\bm{n}}$ is consistent with vanishing $\bm{c}_{\bm{q}}$ means that mixing equilibrium, which is defined by the condition that the $\bm{n} = \bm{0}$ vectors generate no motion [Eq.~\eqref{eq:mixeqHP}], can be static in the presence of flavor waves. In particular, the absence of secular driving of $\bm{P}^{\textrm{eq}}_{\bm{q}}$ by $\bm{c}_{\bm{q}}$ would be ensured if at each $\bm{n}$ there were a fully synchronized configuration of flavor-wave polarizations $\bm{P}_{\bm{q},\bm{n}}$.

In this work we posit $\bm{c}_{\bm{q}}$ to be related to the astrophysical driving, the idea being that subgrid correlations arise from the response of neutrinos to being driven away from equilibrium. Roughly speaking, we anticipate 
\begin{equation}
    \bm{c}_{\bm{q}} \sim \mathcal{O}(l^{-1}_{\textrm{astro}}).
\end{equation}
The influence of $\bm{c}_{\bm{q}}$ is therefore negligible on subgrid scales. Consequently $|\langle \bm{P}_{\bm{q}}\rangle|$ and $\epsilon_{\bm{q}}$ are approximately constant and should indeed be fixed for the purpose of determining local mixing equilibrium. In other words, these quantities emerge as approximate invariants as a result of the system maintaining a flavor configuration close to mixing equilibrium, with the smallness of $\bm{c}_{\bm{q}}$ being due to the slowness of the astrophysical driving.

We are excluding the possibility that $\bm{c}_{\bm{q}} \gg \mathcal{O}(l^{-1}_{\textrm{astro}})$. In this case, $|\langle \bm{P}_{\bm{q}}\rangle|$ and therefore $\bm{P}_{\bm{q}}^{\textrm{eq}}$ would change on subgrid scales. $\bm{c}_{\bm{q}}$ is closely related to flavor instabilities. Strong instabilities 
are precluded in miscidynamics because the system evolves near stable mixing equilibrium. Instabilities, such as those due to weak angular crossings \cite{fiorillo2024fast, liu2024quasi, xiong2025robust}, can only manifest at grid level. In rough terms, strong and weak instabilities correspond to $\bm{c}_{\bm{q}} \gg \mathcal{O}(l^{-1}_{\textrm{astro}})$ and $\bm{c}_{\bm{q}} \sim \mathcal{O}(l^{-1}_{\textrm{astro}})$, respectively.

\subsection{Effective disappearance of fine-grained oscillation invariants}

In the collisionless limit, polarization magnitudes are conserved along neutrino trajectories:
\begin{equation}
    |\bm{P}_{\bm{q}}(t, \bm{r}_0+\bm{\hat{q}}t)| = \textrm{constant}.
\end{equation}
This may suggest that $|\bm{P}_{\bm{q}}(t, \bm{r}_0+\bm{\hat{q}}t)|$ should be regarded as an oscillation invariant as well. Unlike $|\langle \bm{P}_{\bm{q}}\rangle|$, it does not emerge as an approximate invariant due to the system being in a certain dynamical regime. It is exactly conserved at the fine-grained level, up to collisional effects.

But on the contrary, we assume $|\bm{P}_{\bm{q}}(t, \bm{r}_0+\bm{\hat{q}}t)|$ to be inaccessible information as far as the coarse-grained dynamics is concerned. Philosophically, this idea is similar to how some isolated systems are able to thermalize to equilibria characterized by macroscopic intensive parameters like temperature, losing dependence on microscopic details despite the fact that Liouville's theorem is satisfied. Here, the fine-grained invariants $|\bm{P}_{\bm{q}}(t, \bm{r}_0+\bm{\hat{q}}t)|$ effectively disappear in the coarse-grained description, allowing for entropy change due to oscillations \cite{johns2023thermodynamics}.

\subsection{Oscillation invariants and spontaneous symmetry breaking}

In some situations $\bm{H}_{\bm{q}}$ has symmetries that permit additional oscillation invariants. For example, consider an isotropic system with Hamiltonian \cite{duan2010collective, hannestad2006self}
\begin{equation}
    \bm{H}_{\omega} = \omega \bm{B} + \mu \bm{D},
\end{equation}
where $\omega = \delta m^2 / 2 |\bm{q}|$ is the vacuum oscillation frequency, $\delta m^2$ is the mass-squared splitting, and $\bm{B} = (\sin 2\theta)\bm{x} - (\cos 2\theta) \bm{z}$ is the mass vector with mixing angle $\theta$. The collective polarization vector is
\begin{equation}
    \bm{D} = \int_0^\infty d\omega ( \bm{P}_{\omega} - \bm{\bar{P}}_{\omega} ).
\end{equation}
Antineutrinos evolve under $\bm{\bar{H}}_{\omega} = \bm{H}_{-\omega}$. The Hamiltonian is rotationally symmetric around $\bm{B}$ when $\bm{\hat{D}} = \pm \bm{B}$. Oscillations conserve $\bm{B}\cdot\bm{D}$.

Spontaneous symmetry breaking occurs if the equilibrium polarizations $\bm{P}^{\textrm{eq}}_{\omega}$ take on values such that $\bm{\hat{D}}^{\textrm{eq}} \neq \pm \bm{B}$. A Goldstone mode emerges in association with the long-wavelength, low-frequency precession of the order parameter $\bm{D}^{\textrm{eq}}$. This fact has already been recognized in studies of spectral swaps, though not expressed in the language we use here. In finding self-consistent adiabatic solutions under a slowly time-varying $\mu(t)$, it is necessary to introduce a slowly changing frequency $\Omega(t)$ with which all $\bm{P}_{\omega}$ precess around $\bm{B}$ \cite{raffelt2007adiabaticity, raffelt2007self2}. The system is at equilibrium only in the rotating frame that removes this co-precession. Mixing equilibrium and $\Omega$ are simultaneously determined by solving the usual self-consistency conditions [Eq.~\eqref{eq:selfcons}] in a rotating frame plus an additional constraint from $\bm{B} \cdot \bm{D}$ conservation.

When applying miscidynamics to systems with additional oscillation invariants, Goldstone modes like $\Omega$ need to be evolved along with the local mixing equilibria. We do not develop this aspect of the theory any further here.

\section{Miscidynamics\label{sec:misc}}

\subsection{Formulation in terms of oscillation invariants}

Miscidynamics is based on the assumption that neutrinos evolve near local mixing equilibrium: $\langle \rho_{\bm{q}} \rangle$ is everywhere close to $\rho^{\textrm{eq}}_{\bm{q}}$. As discussed in the previous section, mixing equilibrium is determined by the local values of the oscillation invariants $\langle P^0_{\bm{q}}\rangle$, $|\langle \bm{P}_{\bm{q}} \rangle|$, and $\epsilon_{\bm{q}}$. Miscidynamics describes the evolution of these coarse-grained quantities.

We assume for simplicity that $\epsilon_{\bm{q}}$ is constant as a function of $t$ and $\bm{r}$, ignoring the possibility that either $\langle \bm{P}_{\bm{q}} \rangle$ or $\langle \bm{H}_{\bm{q}}\rangle$ passes through zero. The miscidynamic equations for $\langle P^0_{\bm{q}}\rangle$ and $|\langle \bm{P}_{\bm{q}} \rangle|$ are derived by imposing local mixing equilibrium on the respective evolution equations:
\begin{gather}
    ( \partial_t + \bm{\hat{q}}\cdot\partial_{\bm{r}}) P^{0,\textrm{eq}}_{\bm{q}} = C^{0,\textrm{eq}}_{\bm{q}}, \notag \\
    ( \partial_t + \bm{\hat{q}}\cdot\partial_{\bm{r}}) |\bm{P}_{\bm{q}}^{\textrm{eq}}| = \bm{\hat{P}}^{\textrm{eq}}_{\bm{q}}\cdot \left( \bm{c}^{\textrm{eq}}_{\bm{q}} + \bm{C}^{\textrm{eq}}_{\bm{q}} \right). \label{eq:miscinv}
\end{gather}
The ``eq'' superscripts indicate that all quantities are in local mixing equilibrium. This condition affects $\bm{C}_{\bm{q}}$ and $C^0_{\bm{q}}$ only inasmuch as they depend on $\bm{P}_{\bm{q}}$. The correlation term is
\begin{equation}
    \bm{c}^{\textrm{eq}}_{\bm{q}} = \langle \bm{H}_{\bm{q}} \times \bm{P}_{\bm{q}} \rangle - \bm{H}^{\textrm{eq}}_{\bm{q}} \times \bm{P}^{\textrm{eq}}_{\bm{q}}. 
\end{equation}
Only the first part contributes nontrivially to Eq.~\eqref{eq:miscinv}.

Suppose that at time $t_i$ we know the values of all of these quantities. Then a large ($\sim t_{\textrm{astro}}$) time step can be taken to $t_{i+1}$ using the equations above. This produces $\langle P^0_{\bm{q}}\rangle$ and $|\langle \bm{P}_{\bm{q}} \rangle|$ at $t_{i+1}$. The subsequent time step to $t_{i+2}$ requires that $\bm{P}_{\bm{q}}^{\textrm{eq}}$ be known. The mixing equilibria at $t_{i+1}$ are determined by solving the self-consistency conditions in Eq.~\eqref{eq:selfcons}.

\subsection{Formulation in terms of density matrices}

To recast the miscidynamic equations for $\langle P^0_{\bm{q}}\rangle$ and $|\langle \bm{P}_{\bm{q}} \rangle|$ as an equation for $\langle\rho_{\bm{q}}\rangle$, consider the frame transformation $U_{\bm{q}} (t, \bm{r})$ defined by
\begin{equation}
    \mathcal{E}_{\bm{q}} = U_{\bm{q}}^\dagger \langle H_{\bm{q}} \rangle U_{\bm{q}},
\end{equation}
where $\mathcal{E}_{\bm{q}}$ is diagonal and $\langle H_{\bm{q}} \rangle$ is the coarse-grained Hamiltonian in the original lab frame. We call this $t$-, $\bm{r}$-, and $\bm{q}$-dependent frame the \textit{local equilibrium frame}. In this frame, the coarse-grained QKE is
\begin{equation}
    i ( \partial_t + \bm{\hat{q}}\cdot\partial_{\bm{r}}) \langle\rho_{\bm{q}}\rangle = [ \mathcal{E}_{\bm{q}} + \mathcal{T}_{\bm{q}}, \langle \rho_{\bm{q}} \rangle] + ic_{\bm{q}} + \langle iC_{\bm{q}} \rangle, \label{eq:cgQKE_new}
\end{equation}
with effective Hamiltonian $\mathcal{E}_{\bm{q}} + \mathcal{T}_{\bm{q}}$ and nonadiabatic part
\begin{equation}
    \mathcal{T}_{\bm{q}} \equiv -i U^\dagger_{\bm{q}}( \partial_t + \bm{\hat{q}}\cdot\partial_{\bm{r}}) U_{\bm{q}}. \label{eq:nonadiabCq}
\end{equation}
Recall that the correlation term in the lab frame is
\begin{equation}
    ic_{\bm{q}} = \sum_{\bm{n}\neq 0} [ H_{\bm{q},-\bm{n}}, \rho_{\bm{q},\bm{n}} ].
\end{equation}
We approximate the collision term using
\begin{equation}
    \langle iC_{\bm{q}} \rangle \cong iC_{\bm{q}} [\langle \rho \rangle].
\end{equation}
The notation on the right-hand side indicates that $iC_{\bm{q}}$, which is a functional of $\rho_{\bm{q}'}$ at all $\bm{q}'$, is evaluated using $\rho_{\bm{q}'} \rightarrow \langle \rho_{\bm{q}'} \rangle$. All of $\mathcal{T}_{\bm{q}}$, $ic_{\bm{q}}$, and $iC_{\bm{q}}$ contribute to nonadiabaticity. We emphasize again that Eq.~\eqref{eq:cgQKE_new} is written in the local equilibrium frame. We explicitly indicate this fact for the Hamiltonian by introducing $\mathcal{E}_{\bm{q}}$ and $\mathcal{T}_{\bm{q}}$, for the purpose of distinguishing the adiabatic and nonadiabatic parts. Although we do not introduce new symbols for the remaining quantities in Eq.~\eqref{eq:cgQKE_new}, they are all evaluated in the local equilibrium frame.

We will be adopting the \textit{quasistatic approximation} throughout this work. Applied to the $\bm{n} = \bm{0}$ evolution, this means that
\begin{equation}
    \mathcal{T}_{\bm{q}} \cong 0. \label{eq:n0QS}
\end{equation}
Later we will apply the quasistatic approximation to the flavor-wave dynamics as well.

Let us split $c_{\bm{q}}$ and $C_{\bm{q}}$ into unitary and non-unitary parts:
\begin{gather}
    c_{\bm{q}} = c_{\bm{q},\textrm{uni}} + c_{\bm{q},\textrm{non}}, \notag \\
    C_{\bm{q}} = C_{\bm{q},\textrm{uni}} + C_{\bm{q},\textrm{non}}.
\end{gather}
The splitting is defined through
\begin{gather}
    \bm{c}_{\bm{q},\textrm{non}} \equiv (\bm{\hat{P}}_{\bm{q}} \cdot \bm{c}_{\bm{q}}) \bm{\hat{P}}_{\bm{q}}, \notag \\
    \bm{c}_{\bm{q},\textrm{uni}} \equiv \bm{c}_{\bm{q}} - \bm{c}_{\bm{q},\textrm{non}}, \notag \\
    c_{\bm{q},\textrm{non}} \equiv \frac{1}{2} (c^0_{\bm{q}} + \bm{c}_{\bm{q},\textrm{non}} \cdot \bm{\sigma} ), \notag \\
    c_{\bm{q},\textrm{uni}} \equiv \frac{1}{2}\bm{c}_{\bm{q},\textrm{uni}} \cdot \bm{\sigma},
\end{gather}
and similarly for $iC_{\bm{q}}$. The unitary and non-unitary labels are motivated by the fact that the instantaneous effect of $ic_{\bm{q},\textrm{uni}}$ and $iC_{\bm{q},\textrm{uni}}$ is to redirect $\bm{P}_{\bm{q}}$ rather than change its magnitude, whereas the instantaneous effect of $ic_{\bm{q},\textrm{non}}$ and $iC_{\bm{q},\textrm{non}}$ is to do the opposite. In the local equilibrium frame, $\bm{P}_{\bm{q}}^{\textrm{eq}}$ remains (anti)aligned with $\bm{\mathcal{E}}^{\textrm{eq}}_{\bm{q}}$, the flavor-space vector defined from $\mathcal{E}_{\bm{q}}^{\textrm{eq}}$. This requires
\begin{equation}
     ic_{\bm{q},\textrm{uni}}^{\textrm{eq}} + iC_{\bm{q},\textrm{uni}}^{\textrm{eq}} = 0 \label{eq:eqframe0}
\end{equation}
in the local equilibrium frame, with the superscripts on $ic^{\textrm{eq}}_{\bm{q},\textrm{uni}}$ and $iC^{\textrm{eq}}_{\bm{q},\textrm{uni}}$ indicating as usual that these quantities are evaluated using local-equilibrium values.

With the above conditions imposed, we obtain the miscidynamic equation for $\langle\rho_{\bm{q}}\rangle$:
\begin{equation}
    i ( \partial_t + \bm{\hat{q}}\cdot\partial_{\bm{r}}) \rho^{\textrm{eq}}_{\bm{q}} = ic^{\textrm{eq}}_{\bm{q},\textrm{non}} + iC^{\textrm{eq}}_{\bm{q},\textrm{non}}. \label{eq:miscrho}
\end{equation}
We stress again that this holds only in the local equilibrium frame. Equation~\eqref{eq:eqframe0}, which must be met if $\rho^{\textrm{eq}}_{\bm{q}}$ is in fact to maintain mixing equilibrium, is a condition on $U_{\bm{q}}$. If the system follows local mixing equilibrium, then there exists a frame transformation $U_{\bm{q}}$ such that Eq.~\eqref{eq:eqframe0} is satisfied, and in this frame the evolution of $\langle \rho_{\bm{q}} \rangle$ is given by Eq.~\eqref{eq:miscrho}.

In writing Eq.~\eqref{eq:miscrho}, we absorb all of the lab-frame, grid-level dynamics of $\bm{\hat{P}}^{\textrm{eq}}_{\bm{q}}$ into $U_{\bm{q}}$. A unitary transformation can always be adopted to absorb the motion of $\bm{\hat{P}}_{\bm{q}}$. But two points are true of local-equilibrium evolution that are not true in general. First, because the oscillation invariants change on the $l_{\textrm{astro}}$ scale, as per Eq.~\eqref{eq:miscinv}, the local equilibrium frame also varies slowly in $t$ and $\bm{r}$. Second, $U_{\bm{q}}$ can be inferred by updating $\langle P^0_{\bm{q}}\rangle$ and $\langle \bm{P}_{\bm{q}} \rangle$ using Eq.~\eqref{eq:miscinv} and then solving the self-consistency conditions. In the general case, by contrast, determining the frame transformation that keeps $\bm{\hat{P}}_{\bm{q}}$ stationary requires solving the full dynamics.

\subsection{Turbulent flavor-wave viscosity}

From Eq.~\eqref{eq:miscrho} we obtain the miscidynamic equation for $\bm{P}_{\bm{q}}$ in the local equilibrium frame:
\begin{equation}
    ( \partial_t + \bm{\hat{q}} \cdot \partial_{\bm{r}} ) \bm{P}^{\textrm{eq}}_{\bm{q}} = -(\gamma_{\bm{q}}  + \Gamma_{\bm{q}})\bm{P}^{\textrm{eq}}_{\bm{q}}. \label{eq:miscvisc}
\end{equation}
Here we are introducing $\gamma_{\bm{q}}$ and $\Gamma_{\bm{q}}$, the depolarization rates due to subgrid correlations and neutrino collisions, respectively:
\begin{align}
    &\gamma_{\bm{q}} \equiv - \frac{\bm{P}^{\textrm{eq}}_{\bm{q}} \cdot \bm{c}^{\textrm{eq}}_{\bm{q},\textrm{non}}}{|\bm{P}^{\textrm{eq}}_{\bm{q}}|^2}, \notag \\
    &\Gamma_{\bm{q}} \equiv - \frac{\bm{P}^{\textrm{eq}}_{\bm{q}} \cdot \bm{C}^{\textrm{eq}}_{\bm{q},\textrm{non}}}{|\bm{P}^{\textrm{eq}}_{\bm{q}}|^2}.
\end{align}
These rates can take on negative values. For example, $\Gamma_{\bm{q}} < 0$ if collisional processes in the medium act to further polarize the neutrino flavor distribution. We use the term \textit{turbulent flavor-wave viscosity} (or \textit{flavor viscosity} for short) to refer to $\gamma_{\bm{q}}$, in analogy with turbulent eddy viscosity in fluids. It can also be regarded as a decay rate into flavor waves, as we show later.

Note that $\gamma_{\bm{q}}$ depends on $\bm{c}_{\bm{q}}$, which in turn depends on subgrid quantities like $\bm{P}_{\bm{q},\bm{n}\neq \bm{0}}$, while Eq.~\eqref{eq:miscvisc} only evolves grid-level ($\bm{n} = \bm{0}$) quantities. This is the closure problem in neutrino quantum kinetics. The simplest closure ignores flavor viscosity, taking $\gamma_{\bm{q}} = 0$. This assumption, when combined with the quasistatic approximation $\mathcal{T}_{\bm{q}} = 0$, leads to adiabatic miscidynamics, with adiabaticity referring to the preservation of $|\langle \bm{P}_{\bm{q}} \rangle|$ by oscillations \cite{johns2023thermodynamics, johns2024subgrid}. Adiabatic miscidynamics, however, makes no distinction between stable and unstable mixing equilibria and therefore has no way of ensuring that the coarse-grained system tracks (perhaps marginally) stable equilibria instead of being pushed into unstable parameter space. In the following sections we develop a more robust closure scheme for $\gamma_{\bm{q}}$.

\section{Weakly coupled flavor waves\label{sec:neareq}}

\subsection{Small coherence}

In this section we consider the QKE in a subgrid region and demonstrate how the \textit{weak-inhomogeneity approximation}---the assumption that the short-time dynamics of $\bm{P}_{\bm{q},\bm{n}\neq \bm{0}}$ is predominantly determined by the $\bm{n} = \bm{0}$ polarization vectors---allows for a generalization of the familiar form of linear analysis based on small flavor coherence. Weak inhomogeneity implies that flavor waves are weakly coupled to one another and feed back weakly on the background polarization. It allows us to treat flavor waves as weakly interacting quasiparticles whose dynamics can be approximated using kinetic theory.

We begin this section by reviewing the standard form of linear analysis \cite{banerjee2011, izaguirre2017}, which we call \textit{small-coherence linear analysis} to distinguish it from the generalization we present next. Consider the QKE in a box with side length $l_{\textrm{box}}$ and volume $V = l_{\textrm{box}}^3$. In small-coherence linear analysis, we write
\begin{equation}
    \rho_{\bm{q}} = \frac{f_{{\nu_e},\bm{q}} + f_{{\nu_x},\bm{q}}}{2} + \frac{f_{{\nu_e},\bm{q}} - f_{{\nu_x},\bm{q}}}{2} \begin{pmatrix}
        s_{\bm{q}} & S_{\bm{q}} \\
        S^*_{\bm{q}} & -s_{\bm{q}}
    \end{pmatrix}
\end{equation}
with $s_{\bm{q}}^2 + |S_{\bm{q}}|^2 = 1$ and assume that 
\begin{equation}
|S_{\bm{q}}| \ll |s_{\bm{q}}| \approx 1.
\end{equation}
The QKE is then linearized in the small parameters $S_{\bm{q}}$, and solutions of the form
\begin{equation}
    S_{\bm{q}} (t, \bm{r}) = \mathcal{S}_{\bm{q},\bm{k}} e^{-i(\Omega t - \bm{k} \cdot \bm{r})}
\end{equation}
are sought, with $\mathcal{S}_{\bm{q},\bm{k}}$ and $\Omega$ being the eigenfunction and eigenvalue associated with a particular solution of the dispersion relation $\mathcal{F}(\Omega,\bm{k}) = 0$. This linearization procedure is based on the assumption of small flavor coherence. Rewritten using the polarization vector, the assumption of small $S_{\bm{q}}$ translates to
\begin{gather}
    |\bm{P}_{\bm{q}}^T| = \sqrt{(P_{\bm{q}}^x)^2 + (P_{\bm{q}}^y)^2} \ll |P_{\bm{q}}^z|.
\end{gather}
In small-coherence linear analysis, every polarization vector is assumed to point nearly along the flavor axis.

There is an obvious generalization of small-coherence linear analysis that is identical except that $S_{\bm{q}}$ represents the coherence in a different basis, or in other words with all $\bm{P}_{\bm{q}}$ nearly (anti)aligned with one another but not necessarily with the flavor axis. The generalization we pursue in the next subsection is more substantive than this one.

\subsection{Weak inhomogeneity}

We no longer assume that coherence is small in any basis. Instead, we adopt the weak-inhomogeneity approximation by discarding quadratic terms in $\bm{P}_{\bm{q},\bm{n}\neq \bm{0}}$. The QKE for $\bm{P}_{\bm{q},\bm{n}\neq\bm{0}}$ is
\begin{align}
    \partial_t \bm{P}_{\bm{q},\bm{n}} = &- (i\bm{\hat{q}} \cdot \bm{k}_{\bm{n}}) \bm{P}_{\bm{q},\bm{n}} \notag \\
    &~~~~~+ \sum_{\bm{m}} \bm{H}_{\bm{q},\bm{n}-\bm{m}} \times \bm{P}_{\bm{q},\bm{m}} + \bm{C}_{\bm{q},\bm{n}}. \label{eq:PqnEOM}
\end{align}
For simplicity of presentation, suppose that $\bm{C}_{\bm{q},\bm{n}} = -\mathcal{D}_{\bm{q}}\bm{P}_{\bm{q},\bm{n}}$
with depolarization rate $\mathcal{D}_{\bm{q}}$. Nonlinear collisional terms could be included, but we neglect them in this paper. Then the linearized evolution is given by
\begin{align}
    \partial_t \bm{P}_{\bm{q},\bm{n}} = &- (i\bm{\hat{q}} \cdot \bm{k}_{\bm{n}} + \mathcal{D}_{\bm{q}}) \bm{P}_{\bm{q},\bm{n}} \notag \\
    &~~~~~+ \bm{H}_{\bm{q},\bm{0}} \times \bm{P}_{\bm{q},\bm{n}} + \bm{H}_{\bm{q},\bm{n}} \times \bm{P}_{\bm{q},\bm{0}}.
\end{align}
Note that this equation is linear in polarization vectors at given $\bm{n} \neq \bm{0}$ but involves neutrinos of all momenta through $\bm{H}_{\bm{q},\bm{0}}$ and $\bm{H}_{\bm{q},\bm{n}}$. The terms in Eq.~\eqref{eq:PqnEOM} sum over $\bm{m}$ that are nonlinear in inhomogeneous polarization vectors have been dropped. The remaining terms are those with $\bm{m} = \bm{0}$ or $\bm{m} = \bm{n}$. As a reminder, we are assuming that all of the subgrid inhomogeneity enters through $\bm{P}_{\bm{q},\bm{n}}$.

Let us introduce the wave function
\begin{equation}
    \Psi_{\bm{n}} \equiv \begin{pmatrix}
        P_{\bm{q}_1,\bm{n}}^x \\
        P_{\bm{q}_1,\bm{n}}^y \\
        P_{\bm{q}_1,\bm{n}}^z \\
        \vdots \\
        P_{\bm{q}_N,\bm{n}}^x \\
        P_{\bm{q}_N,\bm{n}}^y \\
        P_{\bm{q}_N,\bm{n}}^z
    \end{pmatrix}. \label{eq:Psin}
\end{equation}
This assumes there are a total of $N$ momenta, written in some arbitrary order. We only show neutrino components explicitly, though of course antineutrinos are easily accommodated. These wave functions have the property $\Psi_{-\bm{n}} = \Psi_{\bm{n}}^*$.

The evolution on a short time scale is
\begin{equation}
    i\partial_t \Psi_{\bm{n}} = \mathcal{H}_{\bm{n}} \Psi_{\bm{n}} \label{eq:PsinEOM}
\end{equation}
with non-Hermitian Hamiltonian
\begin{widetext}
\begin{equation}
    \mathcal{H}_{\bm{n}} = \begin{pmatrix}
        \bm{\hat{q}}_1\cdot\bm{k}_{\bm{n}} -i \mathcal{D}_{\bm{q}_1} & -i\langle H_{\bm{q}_1}^z \rangle & i\langle H_{\bm{q}_1}^y \rangle & & & & 0 & i \mu f_{\bm{q}_1,\bm{q}_N}\langle P_{\bm{q}_1}^z \rangle & -i \mu f_{\bm{q}_1,\bm{q}_N} \langle P_{\bm{q}_1}^y \rangle \\
        i \langle H_{\bm{q}_1}^z \rangle & \bm{\hat{q}}_1\cdot\bm{k}_{\bm{n}} -i \mathcal{D}_{\bm{q}_1} & -i \langle H_{\bm{q}_1}^x \rangle & & \cdots & & -i \mu f_{\bm{q}_1,\bm{q}_N} \langle P_{\bm{q}_1}^z \rangle & 0 & i \mu f_{\bm{q}_1,\bm{q}_N} \langle P_{\bm{q}_1}^x \rangle  \\
        -i \langle H_{\bm{q}_1}^y \rangle & i \langle H_{\bm{q}_1}^x \rangle & \bm{\hat{q}}_1\cdot\bm{k}_{\bm{n}} -i \mathcal{D}_{\bm{q}_1} & & & & i \mu f_{\bm{q}_1,\bm{q}_N} \langle P_{\bm{q}_1}^y \rangle & -i \mu f_{\bm{q}_1,\bm{q}_N} \langle P_{\bm{q}_1}^x \rangle & 0 \\
        & & & & & & & & \\
        & \vdots & & & \ddots & & & \vdots & \\
        & & & & & & & & \\
        0 & i \mu f_{\bm{q}_N,\bm{q}_1} \langle P_{\bm{q}_N}^z \rangle & -i \mu f_{\bm{q}_N,\bm{q}_1} \langle P_{\bm{q}_N}^y \rangle & & & & \bm{\hat{q}}_N\cdot\bm{k}_{\bm{n}} -i \mathcal{D}_{\bm{q}_N} & -i \langle H_{\bm{q}_N}^z \rangle & i \langle H_{\bm{q}_N}^y \rangle \\
        -i \mu f_{\bm{q}_N,\bm{q}_1}\langle P_{\bm{q}_N}^z \rangle & 0 & i \mu f_{\bm{q}_N,\bm{q}_1} \langle P_{\bm{q}_N}^x \rangle & & \cdots & & i \langle H_{\bm{q}_N}^z \rangle & \bm{\hat{q}}_N\cdot\bm{k}_{\bm{n}} -i \mathcal{D}_{\bm{q}_N} & -i \langle H_{\bm{q}_N}^x \rangle \\
        i \mu f_{\bm{q}_N,\bm{q}_1} \langle P_{\bm{q}_N}^y \rangle & -i \mu f_{\bm{q}_N,\bm{q}_1} \langle P_{\bm{q}_N}^x \rangle & 0 & & & & -i \langle H_{\bm{q}_N}^y \rangle & i \langle H_{\bm{q}_N}^x \rangle & \bm{\hat{q}}_N\cdot\bm{k}_{\bm{n}} -i \mathcal{D}_{\bm{q}_N}
    \end{pmatrix}\label{eq:HnFW}
\end{equation}
\end{widetext}
using the notation
\begin{equation}
    f_{\bm{q}_i,\bm{q}_j} \equiv 1 - \bm{\hat{q}}_i \cdot \bm{\hat{q}}_j.
\end{equation}
The flavor-wave Hamiltonians have the property
\begin{equation}
    \mathcal{H}_{-\bm{n}} = -\mathcal{H}_{\bm{n}}^*. \label{eq:Hminusn}
\end{equation}
Let $\psi_{\bm{n},i}$ be an eigenvector of $\mathcal{H}_{\bm{n}}$ with eigenvalue $\Omega_{\bm{n},i}$. Then, by Eq.~\eqref{eq:Hminusn}, $\psi_{\bm{n},i}^*$ is an eigenvector of $\mathcal{H}_{-\bm{n}}$ with eigenvalue $-\Omega_{\bm{n},i}^*$.

Weak-inhomogeneity linear analysis is performed by solving the $\mathcal{H}_{\bm{n}}$ eigensystem. It generalizes small-coherence linear analysis in several respects. First, the $\bm{n} = \bm{0}$ polarization vectors do not need to be nearly (anti)aligned with one another. They could, for instance, be in any mixing-equilibrium configuration.

Second, the weak-inhomogeneity approximation allows us to determine eigenvectors with complete flavor-space information: full polarization vectors $\bm{P}_{\bm{q},\bm{n}}$, not just coherences $S_{\bm{q},\bm{k}}$. 

Third, the weak-inhomogeneity approximation actually allows for greater inhomogeneity than the small-coherence approximation does. We require that $|\bm{P}_{\bm{q},\bm{n}}|$ be small for every $\bm{n}$, and therefore that $|\bm{P}_{\bm{q},\bm{n}}^T|$, the part transverse to the flavor axis, also be small. But $|\bm{P}_{\bm{q}}^T (t,\bm{r})|$ does not need to be small at every $\bm{r}$ as it does under the small-coherence approximation. Weak inhomogeneity allows for large inhomogeneous fluctuations as a function of $\bm{r}$ as long as nonlinear flavor-wave couplings and their backreaction on the background polarizations remain negligible.

As a final note, we remark that weak-inhomogeneity linear analysis can easily be extended to three flavors if an 8-dimensional polarization vector $\bm{P}_{\bm{q}}$ is defined using Gell-Mann matrices and $\Psi_{\bm{n}}$ is redefined as a vector of dimension $8N$, with the components of $\bm{P}_{\bm{q},\bm{n}}$ stacked in the obvious generalization of Eq.~\eqref{eq:Psin}.

\subsection{Flavor-wave density matrices\label{sec:FWdm}}

We can put the flavor waves on the same footing as the neutrino density matrices by defining the matrix $R_{\bm{n}\bm{m}}$ with components
\begin{equation}
    [R_{\bm{n}\bm{m}} (t)]_{ij} = \Psi_{\bm{m},i} \Psi^\dagger_{\bm{n},j},
\end{equation}
where $\Psi_{\bm{m},i}$ is the $i$th component of $\Psi_{\bm{m}}$.

To make the parallel with $\rho$ even closer, consider an infinite medium and define the flavor-wave Wigner function $R$ through
\begin{align}
    &[R (t, \bm{r}, \bm{k})]_{ij} = \notag \\
    &~~~~~~\int \frac{d^3 \bm{\delta k}}{(2\pi)^3} e^{i \bm{\delta k} \cdot \bm{r}} \Psi^\dagger_j (t, \bm{k} - \bm{\delta k} / 2 ) \Psi_i (t, \bm{k} + \bm{\delta k} / 2 ). \label{eq:FWwigner}
\end{align}
Here $\bm{r}$ and $\bm{k}$ are the position and momentum coordinates in the flavor-wave phase space. It is customary to call $\rho$ a neutrino density matrix even though it is actually a Wigner function \cite{sigl1993general, vlasenko2014neutrino, stirner2018liouville},
\begin{align}
    &[\rho_{\bm{q}}(t,\bm{r})]_{ij} = \notag \\
    &~~~~~~\int \frac{d^3 \bm{\delta q}}{(2\pi)^3} e^{i \bm{\delta q} \cdot \bm{r}} \left\langle a_j^\dagger (t, \bm{q} - \bm{\delta q} / 2 ) a_i (t, \bm{q} + \bm{\delta q} / 2 ) \right\rangle.
\end{align}
By stretching nomenclature in the same way, let us refer to $R$ as a \textit{flavor-wave density matrix}. Note, however, the distinction in the meanings of the $ij$ indices on $R$ and $\rho$. The indices on $R$ are mixed momentum--flavor labels, reflecting the fact that $\Psi_{\bm{n}}$ incorporates collectivity in $\bm{q}$. We will use $R$ instead of $\Psi_{\bm{n}}$ in the next section's description of flavor-wave transport on slowly varying backgrounds.

\section{Flavor-wave dynamics\label{sec:fwdynamics}}

\subsection{Static background}

The usual motivation for small-coherence linear analysis is to assess the stability of solutions to neutrino classical kinetics. Our motivation here is different. We are not interested in assessing stability per se. Rather, our aim is to develop approximations of the flavor-wave dynamics and thereby of the flavor viscosity $\gamma_{\bm{q}}$. In this subsection we consider flavor-wave dynamics under the assumption that $\mathcal{H}_{\bm{n}}$ is time-independent. Later in this section we will introduce temporal and spatial variation on a macroscopic scale. We also assume here that $\mathcal{H}_{\bm{n}}$ is diagonalizable, so that the evolution of $\Psi_{\bm{n}}$ can be written in terms of an eigenmode expansion. In particular we assume that the system is away from any exceptional points, where the eigensystem becomes defective \cite{heiss2012physics}.

The decomposition of $\Psi_{\bm{n}}$ into flavor-wave eigenmodes is
\begin{equation}
    \Psi_{\bm{n}}(t) = \sum_{i=1}^{3N} A_{\bm{n},i}(t) \psi_{\bm{n},i} e^{-i \Omega_{\bm{n},i}t}\label{eq:Psintexpansion}
\end{equation}
with complex coefficient $A_{\bm{n},i}$ and eigenvector $\psi_{\bm{n},i}$. The eigenstates are on-shell flavor waves, the emergent quasiparticles of the neutrino flavor field. They possess ``energy'' given by eigenvalue real part $\Omega^{\textrm{R}}_{\bm{n},i} \equiv \textrm{Re}(\Omega_{\bm{n},i})$. The decay/growth rate $\Omega^{\textrm{I}}_{\bm{n},i} \equiv \textrm{Im}(\Omega_{\bm{n},i})$ is explored later. The index $i$ is analogous to a band label in condensed matter, so that $\Omega^{\textrm{R}}_{\bm{n},i}$ represents the band structure of a neutrino system. Some $\psi_{\bm{n},i}$ may be single-$\bm{q}$ excitations while others are collective excitations in momentum.

The flavor-wave equation of motion is
\begin{equation}
    i \partial_t \Psi_{\bm{n}} = \mathcal{H}_{\bm{n}} \Psi_{\bm{n}} + N_{\bm{n}}. \label{eq:PsinEOMNL}
\end{equation}
We have restored the nonlinear term $N_{\bm{n}}$ but postpone a detailed discussion of it until the next subsection. Let $S_{\bm{n}}$ be the (generally non-unitary) matrix that diagonalizes $\mathcal{H}_{\bm{n}}$:
\begin{equation}
    S_{\bm{n}}^{-1} \mathcal{H}_{\bm{n}} S_{\bm{n}} = \Omega_{\bm{n}},\label{eq:Sndiagonalizes}
\end{equation}
where $\Omega_{\bm{n}}$ is a diagonal matrix with elements $[\Omega_{\bm{n}}]_{ii} = \Omega_{\bm{n},i}$. Moving into the flavor-wave propagation (instantaneous-eigenstate) basis, $\Psi_{\bm{n}}$  transforms into
\begin{equation}
    \Phi_{\bm{n}} = S_{\bm{n}}^{-1} \Psi_{\bm{n}},
\end{equation}
and from Eq.~\eqref{eq:PsinEOMNL} we obtain
\begin{equation}
    i\partial_t \Phi_{\bm{n}} = \Omega_{\bm{n}}\Phi_{\bm{n}} + S_{\bm{n}}^{-1} N_{\bm{n}}. \label{eq:PhinEOMNL}
\end{equation}

The eigenmode expansion in the propagation basis is
\begin{equation}
    \Phi_{\bm{n}}(t) = \sum_{i=1}^{3N} A_{\bm{n},i}(t) \phi_{\bm{n},i} e^{-i\Omega_{\bm{n},i}t},
\end{equation}
with right eigenmodes $\phi_{\bm{n},i} = S^{-1}_{\bm{n}}\psi_{\bm{n},i}$. We also introduce the left eigenmodes $\hat{\phi}_{\bm{n},i}$. The right and left eigenmodes are normalized such that they satisfy the biorthogonality relation
\begin{equation}
    \hat{\phi}_{\bm{n},j}^* \phi_{\bm{n},i} = \delta_{j,i}.\label{eq:phibiortho}
\end{equation}
The overlap with a particular eigenmode is
\begin{align}
    \Phi_{\bm{n},i}
    \equiv \hat{\phi}_{\bm{n},i}^* \Phi_{\bm{n}}
    &= A_{\bm{n},i} e^{-i\Omega_{\bm{n},i}t} \notag \\
    &\equiv |\Phi_{\bm{n},i}|e^{i\varphi_{\bm{n},i}}.
    \label{eq:varphidef}
\end{align}
where in the second line we introduce the real phase $\varphi_{\bm{n},i}$. Multiplying Eq.~\eqref{eq:PhinEOMNL} on the left by $\hat{\phi}_{\bm{n},i}^*$ produces
\begin{equation}
    i \partial_t \Phi_{\bm{n},i} = \Omega_{\bm{n},i}\Phi_{\bm{n},i} + \left( S_{\bm{n}}^{-1} N_{\bm{n}} \right)_i. \label{eq:PhiniEOMNL}
\end{equation}
Taking real and imaginary parts of Eq.~\eqref{eq:PhiniEOMNL}, we obtain, respectively, the phase evolution equation
\begin{equation}
    \frac{d\varphi_{\bm{n},i}}{dt}
    = -\Omega^{\textrm{R}}_{\bm{n},i}
    -\frac{\mathcal{N}^{\textrm{R}}_{\bm{n},i}}
    {|\Phi_{\bm{n},i}|},
    \label{eq:phaseeom}
\end{equation}
and the amplitude evolution equation
\begin{equation}
    \frac{d|\Phi_{\bm{n},i}|}{dt}
    = \Omega^{\textrm{I}}_{\bm{n},i}|\Phi_{\bm{n},i}|
    + \mathcal{N}^{\textrm{I}}_{\bm{n},i}.
    \label{eq:ampeom}
\end{equation}
where the original nonlinear term enters via
\begin{equation}
    \mathcal{N}_{\bm{n},i} \equiv \left( S_{\bm{n}}^{-1} N_{\bm{n}} \right)_i e^{-i \varphi_{\bm{n},i}}.\label{eq:mathcalNdefn}
\end{equation}
We next discuss the kinetic approximation of $\mathcal{N}_{\bm{n},i}$ and flavor viscosity $\gamma_{\bm{q}}$.

\subsection{Flavor-wave kinetics\label{sec:fwkinetics}}

We first focus on the nonlinear term for $\bm{n} \neq \bm{0}$, treating $\bm{n} = \bm{0}$ separately later in this subsection.

The structure of $N_{\bm{n}}$ is implied by the equation of motion of $\bm{P}_{\bm{q},\bm{n}}$ [Eq.~\eqref{eq:PqnEOM}]. In particular, $N_{\bm{n}}$ encodes the sum
\begin{equation}
    \sum_{\bm{m}\notin\lbrace \bm{0}, \bm{n} \rbrace} \bm{H}_{\bm{q},\bm{n}-\bm{m}} \times \bm{P}_{\bm{q},\bm{m}},
\end{equation}
with the $\bm{m} = \bm{0}, \bm{n}$ terms excluded because they are contained in $\mathcal{H}_{\bm{n}}$. The components are
\begin{equation}
    N_{\bm{n},i} = \sum_{\bm{m} \notin \lbrace \bm{0}, \bm{n}\rbrace} \Psi_{\bm{n}-\bm{m}}^T iV_i \Psi_{\bm{m}}, \label{eq:Nni}
\end{equation}
where $V_i$ is a $3N \times 3N$ flavor-wave coupling matrix. As an illustration of the relationship between the coupling matrices and polarization vectors, consider the equation of motion
\begin{align}
    \partial_t{P}_{\bm{q}_i,\bm{n}}^x &= \partial_t\Psi_{\bm{n},3(i-1)+1} \notag \\
    &= -i (\mathcal{H}_{\bm{n}}\Psi_{\bm{n}})_{3(i-1)+1} \notag \\
    &~~~~~~~~~+ \sum_{\bm{m} \notin \lbrace \bm{0}, \bm{n} \rbrace} \Psi_{\bm{n}-\bm{m}}^T V_{3(i-1)+1} \Psi_{\bm{m}}.
\end{align}
Coupling matrix $V_{3(i-1)+1}$ appears in the equation for the $x$ component of the polarization vector for momentum $\bm{q}_i$. Written out explicitly,
\begin{equation}
    V_{3(i-1)+1} = \mu \begin{pmatrix}
        && 0 & 0 & 0 && \\
        & \cdots & 0 & 0 & f_{\bm{q}_i,\bm{q}_1} & \cdots & \\
        && 0 & -f_{\bm{q}_i,\bm{q}_1} & 0 && \\
        &&  & \vdots &  && \\
        && 0 & 0 & 0 && \\
        & \cdots & 0 & 0 & f_{\bm{q}_i,\bm{q}_N} & \cdots & \\
        && 0 & -f_{\bm{q}_i,\bm{q}_N} & 0 &&
    \end{pmatrix}, \label{eq:Vmatx}
\end{equation}
where columns $3(i-1)+1$ through $3(i-1)+3$ are explicitly shown. The $3 \times 3$ blocks in Eq.~\eqref{eq:Vmatx} correspond to couplings to neutrinos with momenta $\bm{q}_1$ and $\bm{q}_N$. For couplings to antineutrinos, the $3 \times 3$ blocks are the same as shown but with $f \rightarrow -f$. Coupling matrices $V_{3(i-1)+2}$ and $V_{3(i-1)+3}$ appear in the equations for the $y$ and $z$ components, respectively. They are
\begin{equation}
    V_{3(i-1)+2} = \mu \begin{pmatrix}
        && 0 & 0 & -f_{\bm{q}_i,\bm{q}_1} && \\
        & \cdots & 0 & 0 & 0 & \cdots & \\
        && f_{\bm{q}_i,\bm{q}_1} & 0 & 0 && \\
        &&  & \vdots &  && \\
        && 0 & 0 & -f_{\bm{q}_i,\bm{q}_N} && \\
        & \cdots & 0 & 0 & 0 & \cdots & \\
        && f_{\bm{q}_i,\bm{q}_N} & 0 & 0 &&
    \end{pmatrix}
\end{equation}
and
\begin{equation}
    V_{3(i-1)+3} = \mu \begin{pmatrix}
        && 0 & f_{\bm{q}_i,\bm{q}_1} & 0 && \\
        & \cdots & -f_{\bm{q}_i,\bm{q}_1} & 0 & 0 & \cdots & \\
        && 0 & 0 & 0 && \\
        &&  & \vdots &  && \\
        && 0 & f_{\bm{q}_i,\bm{q}_N} & 0 && \\
        & \cdots & -f_{\bm{q}_i,\bm{q}_N} & 0 & 0 & \cdots & \\
        && 0 & 0 & 0 &&
    \end{pmatrix},\label{eq:Vmatz}
\end{equation}
displaying the same columns as before and again assuming couplings to neutrinos as opposed to antineutrinos.

We expand $\Psi$ in eigenmodes of $\mathcal{H}$ and use Eq.~\eqref{eq:varphidef} to write
\begin{align}
    N_{\bm{n},i} = \sum_{\bm{m} \notin \lbrace \bm{0}, \bm{n}\rbrace} \sum_{j,l} &|\Phi_{\bm{n}-\bm{m},j}| |\Phi_{\bm{m},l}| e^{i(\varphi_{\bm{n}-\bm{m},j}+\varphi_{\bm{m},l})} \notag \\
    &\times \psi_{\bm{n}-\bm{m},j}^T i V_i \psi_{\bm{m},l}.\label{eq:Nniexpanded}
\end{align}
Into this we insert the short-time approximation
\begin{equation}
    \varphi_{\bm{n},i} \approx \varphi^0_{\bm{n},i} - \Omega^{\textrm{R}}_{\bm{n},i} t,\label{eq:varphiShortTime}
\end{equation}
where $\varphi^0_{\bm{n},i} \equiv \arg(\Phi_{\bm{n},i}(t_0))$ at reference time $t_0 = 0$. Here we are appealing to the smallness of $\mathcal{N}/|\Phi|$ and $\Omega^{\textrm{I}}$ relative to $\Omega^{\textrm{R}}$. We then see that $\mathcal{N}_{\bm{n},i}$ contains factors of
\begin{align}
    &e^{i(\varphi_{\bm{n}-\bm{m},j}+\varphi_{\bm{m},l}-\varphi_{\bm{n},i})} \approx \notag \\
    & ~~~~~~ e^{i(\varphi^0_{\bm{n}-\bm{m},j}+\varphi^0_{\bm{m},l}-\varphi^0_{\bm{n},i})}e^{-i (\Omega^{\textrm{R}}_{\bm{n}-\bm{m},j} + \Omega^{\textrm{R}}_{\bm{m},l} - \Omega^{\textrm{R}}_{\bm{n},i})t}.\label{eq:expvarphiShortTime}
\end{align}
Terms in $\mathcal{N}_{\bm{n},i}$ rapidly oscillate and average out to zero unless the phase is stationary:
\begin{equation}
    \Omega^{\textrm{R}}_{\bm{n}-\bm{m},j} + \Omega^{\textrm{R}}_{\bm{m},l} - \Omega^{\textrm{R}}_{\bm{n},i} = 0. \label{eq:rescond}
\end{equation}
We therefore adopt the \textit{secular approximation} by making the replacement
\begin{align}
    &e^{i(\varphi_{\bm{n}-\bm{m},j}+\varphi_{\bm{m},l}-\varphi_{\bm{n},i})} \longrightarrow \notag \\
    & ~~~~~~ e^{i(\varphi_{\bm{n}-\bm{m},j}+\varphi_{\bm{m},l}-\varphi_{\bm{n},i})}
    \delta_{\Omega^{\textrm{R}}_{\bm{n}-\bm{m},j} + \Omega^{\textrm{R}}_{\bm{m},l}, \Omega^{\textrm{R}}_{\bm{n},i}}. \label{eq:secular}
\end{align}
Note that we do not use $\varphi^0$ on the right-hand side. In this way we allow for relative phases to slowly evolve under $\mathcal{N}$ but perpetually neglect the terms that rapidly average out.

Applying Eq.~\eqref{eq:secular} to $\mathcal{N}_{\bm{n},i}$ on the right-hand sides of Eqs.~\eqref{eq:phaseeom} and \eqref{eq:ampeom} produces approximate equations of motion for $\varphi_{\bm{n},i}$ and $|\Phi_{\bm{n},i}|$. By using the secular approximation, we eliminate the rapidly varying three-wave coupling terms. Relative phases are still consequential and are ultimately related back to initial conditions.

Now let us consider the nonlinear terms in the $\bm{n} = \bm{0}$ equations of motion. Here we are concerned with the flavor viscosity
\begin{equation}
    \gamma_{\bm{q}_i} = -\frac{\bm{\hat{P}}_{\bm{q}_i,\bm{0}}}{|\bm{P}_{\bm{q}_i,\bm{0}}|} \cdot \sum_{\bm{n}\neq\bm{0}} \bm{H}_{\bm{q}_i,-\bm{n}} \times \bm{P}_{\bm{q}_i,\bm{n}}.
\end{equation}
Recasting this equation in terms of $\Psi$ and $V$, we obtain
\begin{align}
    \gamma_{\bm{q}_i} = &-\frac{1}{|\bm{P}_{\bm{q}_i,\bm{0}}|^2}\sum_{\bm{n} \neq \bm{0}} \Psi^T_{-\bm{n}} \big( P^x_{\bm{q}_i,\bm{0}} V_{3(i-1)+1} \notag \\
    &~~~~~~~+ P^y_{\bm{q}_i,\bm{0}} V_{3(i-1)+2} + P^z_{\bm{q}_i,\bm{0}} V_{3(i-1)+3} \big)\Psi_{\bm{n}}. \label{eq:gammaqsec}
\end{align}
From here we can expand $\Psi$ in eigenmodes and apply the secular approximation as before. The difference is that the condition for the retained terms is
\begin{equation}
    \Omega^{\textrm{R}}_{-\bm{n},j} + \Omega^{\textrm{R}}_{\bm{n},i} = 0. \label{eq:rescond0}
\end{equation}
Recalling the text below Eq.~\eqref{eq:Hminusn}, we note that if $\Omega_{\bm{n},i}$ is an eigenvalue of $\mathcal{H}_{\bm{n}}$, then $-\Omega^*_{\bm{n},i}$ is an eigenvalue of $\mathcal{H}_{-\bm{n}}$. We can choose the $i$ labels such that
\begin{equation}
    \Omega_{-\bm{n}, i} = -\Omega_{\bm{n},i}^*. \label{eq:Omegamn}
\end{equation}
Therefore, under the secular approximation, even the relative phases drop out of $\gamma_{\bm{q}}$ except in cases of real-part degeneracies at the same $\bm{n}$: $\Omega^{\textrm{R}}_{\bm{n},i} = \Omega^{\textrm{R}}_{\bm{n},j}$. 

The secular approximation isolates the \textit{resonant interactions} involving neutrinos ($\langle \bm{P}_{\bm{q}} \rangle$) and flavor waves ($\Phi_{\bm{n},i}$). The resonance conditions in Eqs.~\eqref{eq:rescond} and \eqref{eq:rescond0} are equivalent to energy conservation in, respectively, three-wave scattering and neutrino decay to two flavor waves (or the inverse process). By inspecting the wave numbers that appear in the resonance conditions, we can see that they also satisfy momentum conservation in the sense that
\begin{equation}
    \bm{k}_{\bm{n}} = \bm{k}_{\bm{n}-\bm{m}} + \bm{k}_{\bm{m}}
\end{equation}
for three-wave scattering and
\begin{equation}
    0 = \bm{k}_{-\bm{n}} + \bm{k}_{\bm{n}}
\end{equation}
for neutrino decay. Momentum conservation is a consequence of homogeneous couplings, not the secular approximation. But under the secular approximation, the nonlinear terms $N_{\bm{n},i}$ (for $\bm{n} \neq \bm{0}$) and $-\gamma_{\bm{q}} \bm{P}^{\textrm{eq}}_{\bm{q}}$ (for $\bm{n} = \bm{0}$) obey the additional constraint of energy conservation and have structures similar to Boltzmann collision integrals.

From Eqs.~\eqref{eq:rescond0} and \eqref{eq:Omegamn} with $i = j$ we can see that neutrino decay into flavor waves is always permitted by energy and momentum conservation. But whether $\langle \bm{P}_{\bm{q}} \rangle$ depolarizes due to neutrino decay also depends on the eigenvector structure through Eq.~\eqref{eq:gammaqsec}. This observation has implications for the degree of flavor thermalization that can be achieved.

In numerical implementations of flavor-wave kinetics, it will be important to ensure that flavor-wave turbulence is not artificially impeded by using a finite number of Fourier modes and, as a result, having sparse or nonexistent resonant interactions.

\subsection{Slowly varying background}

We now introduce slow variation in the background on which the flavor waves are defined. We will adopt the \textit{ray} (or \textit{geometric-optics}) \textit{approximation} in which flavor waves propagate with group velocities and forces implied by weak-inhomogeneity linear analysis. The geometric-optics treatment of flavor-wave propagation is comparable to the quantum-kinetic treatment of neutrino propagation. Semiclassical ($\partial_t + \bm{\hat{q}}\cdot\partial_{\bm{r}}$) propagation in the QKE arises from a gradient expansion applied to the Wigner function $\rho_{\bm{q}}$ \cite{sigl1993general, vlasenko2014neutrino, stirner2018liouville}. In Eq.~\eqref{eq:FWwigner} we introduced the Wigner function $R$ for flavor waves. Below we apply a short-wavelength analysis to the evolution of $R$ to derive the ``semiclassical'' flavor-wave Liouville derivative.

Thus far we have considered $\Psi_{\bm{n}}(t)$ defined within an artificially isolated region. To adapt our analysis to the global problem, we return to our separation of the polarization field into the local mean and fluctuation,
\begin{equation}
    \bm{P}_{\bm{q}}(t,\bm{r}) =
    \langle\bm{P}_{\bm{q}}\rangle(t,\bm{r})
    + \delta\bm{P}_{\bm{q}}(t,\bm{r}),
    \label{eq:fwmeansplit}
\end{equation}
where $\langle\bm{P}_{\bm{q}}\rangle \cong \bm{P}^{\textrm{eq}}_{\bm{q}}$ is determined by miscidynamics [Eqs.~\eqref{eq:miscinv} and \eqref{eq:selfcons}]. As in previous sections, $\langle\cdots\rangle$ denotes an average over regions that are larger than the flavor-wave wavelengths but smaller than the scale on which the local mixing equilibrium varies.

We introduce a $3N$-component field by stacking the polarization fluctuations,
\begin{equation}
    \Psi(t,\bm{r}) \equiv
    \begin{pmatrix}
        \delta\bm{P}_{\bm{q}_1}(t,\bm{r}) \\
        \vdots \\
        \delta\bm{P}_{\bm{q}_N}(t,\bm{r})
    \end{pmatrix}.
    \label{eq:fwrealspace}
\end{equation}
In a periodic box, the $\bm{n}\neq\bm{0}$ spatial Fourier coefficients of $\Psi (t, \bm{r})$ agree with $\Psi_{\bm{n}}$ in Eq.~\eqref{eq:Psin}. The Wigner function
\begin{align}
    &R(t,\bm{r},\bm{k}) = \notag\\
    &\hspace{.3 in}\int d^3\bm{\delta r}\,
    e^{-i\bm{k}\cdot\bm{\delta r}}
    \Psi(t,\bm{r}+\bm{\delta r}/2)
    \Psi^\dagger(t,\bm{r}-\bm{\delta r}/2)
    \label{eq:fwWignerReal}
\end{align}
is equivalent to the flavor-wave density matrix defined in Eq.~\eqref{eq:FWwigner}. Note that $R$ is Hermitian and satisfies
\begin{equation}
    \int\frac{d^3\bm{k}}{(2\pi)^3}R(t,\bm{r},\bm{k})
    =\Psi(t,\bm{r})\Psi^\dagger(t,\bm{r}).
    \label{eq:fwWignerMarginal}
\end{equation}
Because the polarization fluctuations are real, $R(t,\bm{r},-\bm{k})=R^*(t,\bm{r},\bm{k})$. The normalization of $R$ follows unambiguously from that of the polarization vectors.

Now we derive the evolution equation for the flavor-wave density matrix directly from the QKE. The equation of motion for the polarization fluctuations is
\begin{align}
    ( \partial_t + \bm{\hat{q}}\cdot\partial_{\bm{r}}) \delta\bm{P}_{\bm{q}} = &\langle \bm{H}_{\bm{q}} \rangle \times \delta \bm{P}_{\bm{q}} + \delta\bm{H}_{\bm{q}}\times\langle\bm{P}_{\bm{q}}\rangle - \mathcal{D}_{\bm{q}}\delta\bm{P}_{\bm{q}} \notag\\
    &+ \delta\bm{H}_{\bm{q}}\times\delta\bm{P}_{\bm{q}} - \bm{c}_{\bm{q}},
\end{align}
using the simplified but illustrative collisional form $\bm{C}_{\bm{q}} = - \mathcal{D}_{\bm{q}}\bm{P}_{\bm{q}}$. Recall that $\bm{c}_{\bm{q}} = \langle \delta\bm{H}_{\bm{q}} \times \delta \bm{P}_{\bm{q}}\rangle$. For $N$ discrete momenta, $\Psi (t, \bm{r})$ is therefore governed by the equation
\begin{equation}
    i\partial_t\Psi  = \hat{\mathcal{H}}\Psi  + N ,
    \label{eq:fwRealEOM}
\end{equation}
where $N (t, \bm{r})$ is quadratic in polarization fluctuations and has components
\begin{widetext}
\begin{equation}
N_{3(i-1)+a}(t,\bm{r})
=
i\mu\sum_{j,b,c}
f_{\bm{q}_i,\bm{q}_j}\epsilon^{abc}
\left[
\delta P_{\bm{q}_j}^{b}(t,\bm{r})
\delta P_{\bm{q}_i}^{c}(t,\bm{r})
-
\langle
\delta P_{\bm{q}_j}^{b}
\delta P_{\bm{q}_i}^{c}
\rangle(t,\bm{r})
\right].\label{eq:N3i1a}
\end{equation}
\end{widetext}
We use a hat to emphasize that $\hat{\mathcal{H}} (t, \bm{r}, -i \partial_{\bm{r}})$ is a matrix-structured real-space differential operator. It is related to $\mathcal{H}_{\bm{n}}$ in Eq.~\eqref{eq:HnFW} by
\begin{equation}
    \hat{\mathcal{H}} (t, \bm{r}, - i \partial_{\bm{r}}) =
    \mathcal{H}_{\bm{n}}(t)|_{\bm{k}_{\bm{n}} \rightarrow - i \partial_{\bm{r}}},
\end{equation}
with the background polarizations appearing in $\mathcal{H}_{\bm{n}}$ evaluated at position $\bm{r}$.

It will be convenient to introduce the notation $X_{\pm} \equiv X (t, \bm{r}_{\pm})$ for vector $X$, with $\bm{r}_{\pm} = \bm{r} \pm \bm{\delta r}/2$, and
\begin{equation}
    \mathcal{W}[X,Y] \equiv \int d^3\bm{\delta r} \, e^{- i \bm{k} \cdot \bm{\delta r}} X_+ Y_-^\dagger
\end{equation}
for a pair of vectors $X$ and $Y$. From the definition of $R$ in Eq.~\eqref{eq:fwWignerReal} and the $\Psi$ equation of motion in Eq.~\eqref{eq:fwRealEOM}, we obtain
\begin{equation}
    i \partial_t R = \mathcal{W}[\hat{\mathcal{H}}\Psi,\Psi] - \mathcal{W}[\Psi,\hat{\mathcal{H}}\Psi] + i \mathcal{C}_R,\label{eq:fwREomWWC}
\end{equation}
where
\begin{equation}
    i \mathcal{C}_R \equiv \mathcal{W}[N,\Psi] - \mathcal{W} [\Psi, N].\label{eq:CRWignerFirst}
\end{equation}
The real-space operator $\hat{\mathcal{H}}$ can be split into separate terms for advection and interactions:
\begin{equation}
    \hat{\mathcal{H}}(t,\bm{r}, -i \partial_{\bm{r}}) = - i \bm{U}\cdot \partial_{\bm{r}} + B(t, \bm{r}),
\end{equation}
where $\bm{U}$ is a vector of matrices whose $a$th Cartesian component is
\begin{equation}
    U_a \equiv \textrm{diag} (\hat{q}_1^a, \hat{q}_1^a, \hat{q}_1^a, \dots, \hat{q}_N^a, \hat{q}_N^a, \hat{q}_N^a)
\end{equation}
and $B(t, \bm{r})$ contains the linearized interactions and damping.

From here the derivation of the linear part of the $R$ equation of motion follows a standard path. See Ref.~\cite{stirner2018liouville} for an example from the neutrino literature. For the advective contribution to $\mathcal{W}[\hat{\mathcal{H}}\Psi, \Psi]$, we find, after some algebra,
\begin{equation}
    \mathcal{W}[-i\bm{U} \cdot \partial_{\bm{r}} \Psi, \Psi] = \bm{U} \cdot\left( \bm{k} - \frac{i}{2} \partial_{\bm{r}}  \right) R.
\end{equation}
For the interaction/damping piece, we use the gradient expansion
\begin{equation}
    B(t,\bm{r}_+) = B(t,\bm{r}) + \frac{\bm{\delta r}}{2} \cdot \partial_{\bm{r}} B(t,\bm{r}) + \dots
\end{equation}
and find
\begin{equation}
    \mathcal{W}[B\Psi, \Psi] = BR + \frac{i}{2}\partial_{\bm{r}}B \cdot \partial_{\bm{k}}R
\end{equation}
up to higher-order gradients. Defining the phase-space Hamiltonian
\begin{equation}
    \mathcal{H} (t, \bm{r}, \bm{k}) = \bm{U} \cdot \bm{k} + B(t, \bm{r})\label{eq:mathcalHPhaseSpace}
\end{equation}
and noting that $\bm{U} = \partial_{\bm{k}}\mathcal{H}$, we combine the contributions above and write
\begin{equation}
    \mathcal{W}[\hat{\mathcal{H}}\Psi,\Psi] = \mathcal{H} R + \frac{i}{2} \left( \partial_{\bm{r}} \mathcal{H} \cdot \partial_{\bm{k}}R  - \partial_{\bm{k}} \mathcal{H} \cdot \partial_{\bm{r}}R \right).
\end{equation}
A similar calculation leads to
\begin{equation}
    \mathcal{W} [\Psi, \hat{\mathcal{H}}\Psi] = R \mathcal{H}^\dagger + \frac{i}{2} \left( \partial_{\bm{r}}R \cdot \partial_{\bm{k}} \mathcal{H}^\dagger -  \partial_{\bm{k}}R \cdot \partial_{\bm{r}}\mathcal{H}^\dagger\right)
\end{equation}
and therefore, from Eq.~\eqref{eq:fwREomWWC},
\begin{widetext}
\begin{equation}
    \partial_tR = -i(\mathcal{H}R-R\mathcal{H}^\dagger)
    +\frac{1}{2} \left[
    \partial_{\bm{r}}\mathcal{H} \cdot \partial_{\bm{k}}R
    +\partial_{\bm{k}}R \cdot \partial_{\bm{r}}\mathcal{H}^\dagger
    -\partial_{\bm{k}}\mathcal{H} \cdot \partial_{\bm{r}}R
    -\partial_{\bm{r}}R \cdot \partial_{\bm{k}}\mathcal{H}^\dagger
    \right] + \mathcal{C}_R,
    \label{eq:fwWignerGradient}
\end{equation}
\end{widetext}
dropping higher-order phase-space derivatives of $B$ and $R$, as is appropriate for slowly varying Wigner functions \cite{sigl1993general}. An equation of motion retaining the higher-order gradients could be written using the Moyal product \cite{stirner2018liouville}. We will return to $\mathcal{C}_R$ below.

Assume that $\mathcal{H}$ is diagonalizable. Let $S(t,\bm{r},\bm{k})$ be the (generally non-unitary) matrix of right eigenvectors of $\mathcal{H}$, with
\begin{equation}
    S^{-1}\mathcal{H}S=\Omega,
    \label{eq:fwLocalDiag}
\end{equation}
where $\Omega_{ij}=\Omega_i\delta_{ij}$. The flavor-wave density matrix in the propagation basis is
\begin{equation}
    F=S^{-1}R(S^{-1})^\dagger.
    \label{eq:fwFdef}
\end{equation}
We use this to relate the derivatives of $F$ to those of $R$. Letting $\alpha$ represent $t$, $\bm{r}$, or $\bm{k}$, we find
\begin{equation}
    \nabla_{\alpha}F = S^{-1} ( \partial_{\alpha} R) (S^{-1})^{\dagger},
    \label{eq:fwCovDef}
\end{equation}
where we have introduced the covariant derivative
\begin{equation}
    \nabla_{\alpha} F \equiv \partial_{\alpha}F + \mathscr{A}_{\alpha}F + F \mathscr{A}_{\alpha}^{\dagger}
\end{equation}
and the connection
\begin{equation}
    \mathscr{A}_{\alpha} \equiv S^{-1}\partial_{\alpha}S.
\end{equation}
From Eq.~\eqref{eq:fwLocalDiag} we also find
\begin{equation}
    \tilde{\nabla}_{\alpha}\Omega = S^{-1} ( \partial_{\alpha}\mathcal{H})S
\end{equation}
in terms of a second covariant derivative
\begin{equation}
    \tilde{\nabla}_{\alpha}\Omega \equiv \partial_{\alpha}\Omega + [\mathscr{A}_{\alpha},\Omega].
\end{equation}
Applying the relations above to Eq.~\eqref{eq:fwWignerGradient} produces
\begin{widetext}
\begin{equation}
    \nabla_t F = -i(\Omega F-F\Omega^\dagger)
    +\frac{1}{2} \left[
    \tilde{\nabla}_{\bm{r}}\Omega \cdot \nabla_{\bm{k}}F
    +\nabla_{\bm{k}}F \cdot (\tilde{\nabla}_{\bm{r}}\Omega)^\dagger
    -\tilde{\nabla}_{\bm{k}}\Omega \cdot \nabla_{\bm{r}}F
    -\nabla_{\bm{r}}F \cdot (\tilde{\nabla}_{\bm{k}}\Omega)^\dagger
    \right] + \mathcal{C}_F,
    \label{eq:fwCovariantTransport}
\end{equation}
\end{widetext}
where $\mathcal{C}_F \equiv S^{-1}\mathcal{C}_R(S^{-1})^\dagger$.

We now impose the quasistatic approximation, neglecting transitions between energy bands driven by the finite rate of change of the background:
\begin{equation}
    \mathscr{A}_{\alpha,ij} \cong \mathscr{A}_{\alpha,ii} \delta_{i,j}.
\end{equation}
Moreover, we assume that the imaginary part $\Omega^{\textrm{I}}$ has small gradients that can be dropped. As a result of these approximations,
\begin{equation}
    \tilde{\nabla}_{\alpha} \Omega \cong \partial_{\alpha}\Omega^{\textrm{R}}.
\end{equation}
Now define the flavor-wave group velocities and forces
\begin{equation}
    \bm{v}^{\textrm{g}}_i \equiv \frac{\partial \Omega_i^{\textrm{R}}}{\partial \bm{k}}, ~~~ \bm{f}_i \equiv -\frac{\partial \Omega_i^{\textrm{R}}}{\partial \bm{r}}\label{eq:groupvelforcedefn}
\end{equation}
along with the averages
\begin{equation}
    \bm{v}^{\textrm{g}}_{ij} \equiv \frac{\bm{v}^{\textrm{g}}_i + \bm{v}^{\textrm{g}}_j}{2}, ~~~ \bm{f}_{ij} \equiv \frac{\bm{f}_i + \bm{f}_j}{2}.\label{eq:groupvelforceavg}
\end{equation}
From Eq.~\eqref{eq:fwCovariantTransport} we obtain
\begin{equation}
    (\nabla_t + \bm{v}^{\textrm{g}}_{ij}\cdot\nabla_{\bm{r}} + \bm{f}_{ij}\cdot\nabla_{\bm{k}})F_{ij} = -i (\Omega_i - \Omega_j^*) F_{ij} + \mathcal{C}_{F,ij}.\label{eq:fwCovariantLiouville}
\end{equation}
The diagonal parts of the connection still appear through
\begin{equation}
    \nabla_{\alpha}F_{ij} \cong \partial_{\alpha}F_{ij} + (\mathscr{A}_{\alpha,ii} + \mathscr{A}_{\alpha,jj}^*) F_{ij}.
\end{equation}
Indeed, we can write the left-hand side of Eq.~\eqref{eq:fwCovariantLiouville} as
\begin{equation}
    \mathcal{L}_{ij}F_{ij} + ( \mathscr{G}_{ij} + \mathscr{G}_{ji}^*)F_{ij}
\end{equation}
using the definitions
\begin{equation}
    \mathcal{L}_{ij} \equiv \partial_t + \bm{v}^{\textrm{g}}_{ij}\cdot \partial_{\bm{r}} + \bm{f}_{ij} \cdot\partial_{\bm{k}}\label{eq:mathcalLdefn}
\end{equation}
and
\begin{equation}
    \mathscr{G}_{ij} \equiv \mathscr{A}_{t,ii} + \bm{v}^{\textrm{g}}_{ij}\cdot \mathscr{A}_{\bm{r},ii} + \bm{f}_{ij} \cdot \mathscr{A}_{\bm{k},ii}.\label{eq:mathscrG}
\end{equation}
Thus Eq.~\eqref{eq:fwCovariantLiouville} is equivalent to
\begin{equation}
    \mathcal{L}_{ij}F_{ij} = -i (\Omega_i - \Omega_j^*) F_{ij} - (\mathscr{G}_{ij} + \mathscr{G}_{ji}^*)F_{ij} + \mathcal{C}_{F,ij}.\label{eq:fwLiouvilleGeometric}
\end{equation}
Examining the left-hand side, we see that we have recovered the expectation from geometric optics that flavor waves propagate through globally inhomogeneous environments in accord with Hamilton's equations
\begin{equation}
    \frac{d\bm{r}}{dt}=\bm{v}^{\textrm{g}},
    \qquad
    \frac{d\bm{k}}{dt}=\bm{f},
    \label{eq:rayapprox}
\end{equation}
albeit with the subtlety that coherences propagate with average group velocities and forces. The $\mathscr{G}$ terms on the right-hand side capture geometric effects related to changes in eigenmodes along phase-space trajectories.

We can simplify this result further for the diagonal element of $F$ because we have the freedom to set
\begin{equation}
    \mathscr{G}_{ii} = 0
\end{equation}
using parallel transport. Thus the quasistatic transport equation for a diagonal element is
\begin{equation}
    (\partial_t + \bm{v}^{\textrm{g}}_{i}\cdot \partial_{\bm{r}} + \bm{f}_{i} \cdot\partial_{\bm{k}})F_{ii} = 2\Omega_i^{\textrm{I}}F_{ii} + \mathcal{C}_{F,ii}.\label{eq:fwDiagonalLiouville}
\end{equation}
The first term on the right-hand side describes linear growth or damping, while the second describes nonlinear wave--wave interactions. Although Eq.~\eqref{eq:fwDiagonalLiouville} depends on the convention of parallel transport, physical flavor-polarization information can always be recovered using $R=SFS^\dagger$.

In the equations above we have made the quasistatic approximation by neglecting transitions between energy levels $\Omega_i$ and $\Omega_j$ induced by the finite rate of change of parameters appearing in $\mathcal{H}$. Although appealing in its simplicity, this approximation is probably inapplicable in many cases. Flavor-wave evolution is unlikely to be quasistatic in scenarios with unstable flavor waves ($\Omega^{\textrm{I}} \neq 0$) due to the appearance of exceptional points in the flavor-wave spectra. When $\Omega_i$ and $\Omega_j$ are degenerate, transitions between the two eigenmodes are unsuppressed and the evolution may be non-quasistatic regardless of how slow the astrophysical driving is. We leave for future work the development of techniques for dealing with non-quasistatic dynamics.

For the nonlinear term, we return to Eq.~\eqref{eq:N3i1a} and rewrite it as
\begin{equation}
    N_i(t,\bm{r}) = \Psi^T iV_i \Psi - \langle \Psi^T iV_i \Psi \rangle,\label{eq:NiReal}
\end{equation}
where in this equation we are using $i \in \lbrace 1, \dots, 3N \rbrace$ to represent a $\Psi$ vector index rather than identify a particular neutrino momentum. On the right-hand side, $V_i$ is one of the coupling matrices from Eqs.~\eqref{eq:Vmatx} through \eqref{eq:Vmatz} and $\Psi$ is a function of $t$ and $\bm{r}$. In a periodic box, the Fourier transform of Eq.~\eqref{eq:NiReal} for $\bm{n}\neq0$ is our earlier result in Eq.~\eqref{eq:Nni} for the nonlinear term $N_{\bm{n}}$ acting on $\Psi_{\bm{n}}$. From Eq.~\eqref{eq:CRWignerFirst} we have
\begin{equation}
    i\mathcal{C}_R = \int d^3\bm{\delta r} e^{-i \bm{k} \cdot \bm{\delta r}} \left( N_+ \Psi_-^{\dagger} - \Psi_+ N_-^{\dagger} \right).
\end{equation}
Combining this with Eq.~\eqref{eq:NiReal}, we see that the nonlinear term acting on $R$ involves a Wigner transform with a three-wave $\Psi \Psi \Psi^*$ structure. Including nonlinear wave--wave interactions in the evolution of the flavor-wave density matrix thus presents a technical challenge. The goal being to pass over to a description of transport in terms of $R$ alone, some kind of closure needs to be prescribed using the methods of wave kinetic theory \cite{newell2011wave, nazarenko2011wave}. Presumably the secular approximation will emerge naturally in this context as a byproduct of using the Markov approximation to take the limits of time integrals to infinity, just like how energy conservation of interactions emerges in quantum kinetic theory \cite{sigl1993general}. We leave the development of flavor-wave collision integrals to future work. 

The nonlinear backreaction of flavor waves on the local mean polarizations $\bm{P}_{\bm{q}}^{\textrm{eq}}$ introduces no additional closure problem besides the one already inherent in the dynamics of $R$. Recall that the miscidynamic equation, expressed in the local equilibrium frame, is
\begin{equation}
    (\partial_t+\bm{\hat q}\cdot\partial_{\bm r})
    \bm{P}_{\bm q}^{\textrm{eq}}
    =-(\gamma_{\bm q}+\Gamma_{\bm q})
    \bm{P}_{\bm q}^{\textrm{eq}},
    \label{eq:fwMiscCoupled}
\end{equation}
where
\begin{equation}
    \gamma_{\bm q} = -\frac{\bm{P}_{\bm q}^{\textrm{eq}}
    \cdot\bm{c}_{\bm q}}{|\bm{P}_{\bm q}^{\textrm{eq}}|^2}.
    \label{eq:fwCgamma}
\end{equation}
In terms of $\Psi$, the flavor-wave viscosity is
\begin{equation}
    \gamma_{\bm{q}_i} = - \frac{1}{| \bm{P}_{\bm{q}_i}^{\textrm{eq}}|^2} \sum_a P_{\bm{q}_i}^{a,\textrm{eq}} \langle \Psi^T V_{3(i-1)+a}\Psi\rangle,
    \label{eq:fwgammaqV}
\end{equation}
which is equivalent to Eq.~\eqref{eq:gammaqsec} in a periodic box. The spatial average $\langle \dots \rangle$ is inherited from $\bm{c}_{\bm{q}}$ and ultimately originates from the splitting in Eq.~\eqref{eq:fwmeansplit} of $\bm{P}_{\bm{q}}(t,\bm{r})$ into a local spatial average and fluctuation. Note that
\begin{equation}
    \Psi^T V_{3(i-1)+a}\Psi = \textrm{Tr} [V_{3(i-1)+a}\Psi\Psi^{\dagger}].
\end{equation}
Combining this with Eq.~\eqref{eq:fwWignerMarginal}, we have
\begin{equation}
    \Psi^T V_{3(i-1)+a}\Psi = \int \frac{d^3\bm{k}}{(2\pi)^3} \textrm{Tr} [V_{3(i-1)+a} R].
\end{equation}
Therefore $\gamma_{\bm{q}_i}$ can be evaluated exclusively in terms of $R$ using the formula
\begin{equation}
    \gamma_{\bm{q}_i} = - \frac{1}{| \bm{P}_{\bm{q}_i}^{\textrm{eq}}|^2} \sum_a P_{\bm{q}_i}^{a,\textrm{eq}} \int \frac{d^3\bm{k}}{(2\pi)^3} \textrm{Tr} [V_{3(i-1)+a} \langle R \rangle],\label{eq:fwgammaqAngleR}
\end{equation}
with $R$ being averaged over the same region in which $\bm{P}_{\bm{q}_i}^{\textrm{eq}}$ is defined. For $R$ having little variation over this region, the angle brackets on the right-hand side can be dropped.

Under Eq.~\eqref{eq:fwgammaqAngleR}, $\gamma_{\bm{q}}$ may exhibit oscillation-scale variation. Rewritten in terms of $F$, Eq.~\eqref{eq:fwgammaqAngleR} reads
\begin{align}
    \gamma_{\bm{q}_i} &= - \frac{1}{| \bm{P}_{\bm{q}_i}^{\textrm{eq}}|^2} \sum_a P_{\bm{q}_i}^{a,\textrm{eq}} \int \frac{d^3\bm{k}}{(2\pi)^3} \notag\\
    &\hspace{1. in}\times\sum_{j,l} \langle [ S^{\dagger} V_{3(i-1)+a} S]_{lj} F_{jl} \rangle,\label{eq:fwgammaqFjl}
\end{align}
with rapid changes in the right-hand side due to the short-time evolution
\begin{equation}
    F_{jl} \sim e^{-i \int^t dt' (\Omega^{\textrm{R}}_j - \Omega^{\textrm{R}}_l)}
\end{equation}
implied by Eq.~\eqref{eq:fwLiouvilleGeometric}. This is the $F$ analogue of the short-time dynamics highlighted in Eqs.~\eqref{eq:varphiShortTime} and \eqref{eq:expvarphiShortTime} as the motivation for applying the secular approximation to the evolution of $\Psi_{\bm{n}}$. We apply the secular approximation here as well, retaining in Eq.~\eqref{eq:fwgammaqFjl} only those terms with
\begin{equation}
    \Omega^{\textrm{R}}_j = \Omega^{\textrm{R}}_l,\label{eq:rescondF0}
\end{equation}
mirroring the resonance condition Eq.~\eqref{eq:rescond0} from the periodic-box analysis. If there are no real-part degeneracies, $\gamma_{\bm{q}}$ depends only on the diagonal elements of $F$.

In summary, we have derived a closed system of equations for flavor-wave transport in slowly varying backgrounds. Unlike in our treatment of flavor-wave dynamics in a periodic box, here we have fallen back on the quasilinear approximation, deferring to future work the development of a closure for flavor-wave kinetics. In the proposal developed in this subsection, flavor-wave density matrices evolve under Eq.~\eqref{eq:fwLiouvilleGeometric} with the nonlinear wave--wave interaction term $\mathcal{C}_F$ set to 0 to escape the closure problem. The local-equilibrium polarization vectors evolve under Eq.~\eqref{eq:fwMiscCoupled} with $\gamma_{\bm q}$ calculated from Eq.~\eqref{eq:fwgammaqFjl} restricted to resonant terms satisfying Eq.~\eqref{eq:rescondF0}.

\section{Thermodynamics\label{sec:thermo}}

\subsection{Entropy and irreversibility}

For this section we assume that $\rho_{\bm{q}}$ is a continuous function of $\bm{q}$. The coarse-grained entropy in a region of volume $V$ is \cite{johns2023thermodynamics}
\begin{align}
    S &= V \int \frac{d^3\bm{q}}{(2\pi)^3} (s_{\bm{q}}+\bar{s}_{\bm{q}}), \label{eq:Sdefn}
\end{align}
with entropy density
\begin{equation}
    s_{\bm{q}} = -\textrm{Tr} \left[ \rho_{\bm{q},\bm{0}} \log \rho_{\bm{q},\bm{0}} + (1-\rho_{\bm{q},\bm{0}}) \log (1-\rho_{\bm{q},\bm{0}})\right] \label{eq:Sqdefn}
\end{equation}
and the same formula for antineutrinos but with $\rho \rightarrow \bar{\rho}$. Plugging the expansion in Pauli matrices [Eq.~\eqref{eq:pauli}] into Eq.~\eqref{eq:Sqdefn}, we obtain
\begin{align}
    s_{\bm{q}} = &- \frac{P^0_{\bm{q},\bm{0}} + |\bm{P}_{\bm{q},\bm{0}}|}{2} \log \frac{P^0_{\bm{q},\bm{0}} + |\bm{P}_{\bm{q},\bm{0}}|}{2} \notag \\
    &- \frac{P^0_{\bm{q},\bm{0}} - |\bm{P}_{\bm{q},\bm{0}}|}{2} \log \frac{P^0_{\bm{q},\bm{0}} - |\bm{P}_{\bm{q},\bm{0}}|}{2} \notag \\
    &- \left( 1 - \frac{P^0_{\bm{q},\bm{0}} + |\bm{P}_{\bm{q},\bm{0}}|}{2} \right) \log \left( 1 - \frac{P^0_{\bm{q},\bm{0}} + |\bm{P}_{\bm{q},\bm{0}}|}{2} \right) \notag \\
    &- \left( 1 - \frac{P^0_{\bm{q},\bm{0}} - |\bm{P}_{\bm{q},\bm{0}}|}{2} \right) \log \left( 1 - \frac{P^0_{\bm{q},\bm{0}} - |\bm{P}_{\bm{q},\bm{0}}|}{2} \right).
\end{align}
$S$ increases as $|\bm{P}_{\bm{q},\bm{0}}|$ decreases, holding all other quantities fixed:
\begin{equation}
    \frac{\delta S}{\delta |\bm{P}_{\bm{q},\bm{0}}|} = -V\tanh^{-1} \left( \frac{2|\bm{P}_{\bm{q},\bm{0}}|}{|\bm{P}_{\bm{q},\bm{0}}|^2 + P^0_{\bm{q},\bm{0}} ( 2 - P^0_{\bm{q},\bm{0}} ) }\right) \leq 0, \label{eq:Sdepol}
\end{equation}
where the inequality comes from $0 \leq |\bm{P}_{\bm{q},\bm{0}}| \leq \textrm{min} (P^0_{\bm{q},\bm{0}}, 2-P^0_{\bm{q},\bm{0}})$ and $0 \leq P^0_{\bm{q},\bm{0}} \leq 2$. In the above we have used the functional derivative
\begin{equation}
    \frac{\delta |\bm{P}_{\bm{q}',\bm{0}}|}{\delta |
    \bm{P}_{\bm{q},\bm{0}}|} = (2\pi)^3
     \delta^3(\bm{q}'-\bm{q}).
\end{equation}
From Eq.~\eqref{eq:Sdepol} it is clear that coarse-grained depolarization equates to entropy growth.

In miscidynamics, flavor viscosity $\gamma_{\bm{q}}$ causes a change in the entropy through the depolarization term $-\gamma_{\bm{q}} \bm{P}^{\textrm{eq}}_{\bm{q}}$ in Eq.~\eqref{eq:miscvisc}. Here we see that $\gamma_{\bm{q}} \neq 0$ results in nonadiabatic evolution in the thermodynamic sense. If $\gamma_{\bm{q}} > 0$, the coarse-grained system heats up as flavor waves are excited. If $\gamma_{\bm{q}} < 0$, flavor waves fuse into neutrinos (in the terminology of flavor-wave kinetics) and the system becomes more ordered.

It is an unsettled question whether an isolated, collisionless neutrino system generically satisfies $dS/dt \geq 0$ as conjectured in Ref.~\cite{johns2023thermodynamics}. (See, however, the very recent affirmative answer given in Ref.~\cite{fiorillo2025dispersion}.) Such behavior signals effective irreversibility in neutrino flavor dynamics.

\subsection{The flavor-temperature hypothesis}

Reference~\cite{johns2023thermodynamics} contemplated the possibility that the coarse-grained flavor distributions maximize $S$ at fixed polarization energy
\begin{align}
    U = \frac{V}{2} \int \frac{d^3\bm{q}}{(2\pi)^3} \bigg( &\big( \bm{H}_{\bm{q},\bm{0}}^{\textrm{v}+\textrm{m}} + \frac{1}{2} \bm{H}_{\bm{q},\bm{0}}^{\nu\nu} \big) \cdot \bm{P}_{\bm{q},\bm{0}} \notag \\
    - &\big( \bm{\bar{H}}_{\bm{q},\bm{0}}^{\textrm{v}+\textrm{m}} + \frac{1}{2} \bm{\bar{H}}_{\bm{q},\bm{0}}^{\nu\nu} \big) \cdot \bm{\bar{P}}_{\bm{q},\bm{0}} \bigg) \label{eq:Udef}
\end{align}
and fixed values of any other coarse-grained invariants. In the above, $\bm{H}^{\textrm{v}+\textrm{m}}_{\bm{q},\bm{0}}$ contains the vacuum and matter terms in $\bm{H}_{\bm{q},\bm{0}}$, and $\bm{H}^{\nu\nu}_{\bm{q},\bm{0}}$ is the neutrino--neutrino forward-scattering part [Eq.~\eqref{eq:Hnunu}]. Note that $U$ includes the energetic contribution only from $\bm{n} = \bm{0}$. Thermalization to this maximum-entropy distribution, if it occurs, is brought about by coherent forward-scattering interactions among neutrinos. We call this possibility the \textit{flavor-temperature hypothesis}.

Let us consider a system with flavor polarizations that are not constrained by any invariants other than $U$. Then we define
\begin{equation}
    S' = S + 2\beta (U - \mathcal{U}),
\end{equation}
where $\beta$ is a Lagrange multiplier and $\mathcal{U}$ is the fixed energy expectation value. We maximize $S'$ with respect to $|\bm{P}_{\bm{q},\bm{0}}|$:
\begin{equation}
    \frac{\delta S}{\delta |\bm{P}_{\bm{q},\bm{0}}|} = -2\beta \frac{\delta U}{\delta |\bm{P}_{\bm{q},\bm{0}}|}.
\end{equation}
From Eq.~\eqref{eq:Udef} we have
\begin{equation}
    \frac{\delta U}{\delta |\bm{P}_{\bm{q},\bm{0}}|} = \frac{V}{2} \bm{H}_{\bm{q},\bm{0}} \cdot \bm{\hat{P}}_{\bm{q},\bm{0}}.
\end{equation}
Using $h_{\bm{q}} \equiv \bm{H}_{\bm{q},\bm{0}}\cdot\bm{\hat{P}}_{\bm{q},\bm{0}}$, we write down the relationship satisfied by the polarization magnitudes in the maximum-entropy state:
\begin{equation}
    \frac{2|\bm{P}_{\bm{q},\bm{0}}|}{|\bm{P}_{\bm{q},\bm{0}}|^2 + P^0_{\bm{q},\bm{0}} ( 2 - P^0_{\bm{q},\bm{0}} ) } = \tanh \beta h_{\bm{q}}. \label{eq:tanh}
\end{equation}
We call $\beta^{-1}$ the \textit{flavor temperature}. It is common to neutrinos of all momenta. The relationship in Eq.~\eqref{eq:tanh} applies to antineutrinos as well but with $\beta \rightarrow -\beta$. We emphasize that, because of the collectivity of $h_{\bm{q}}$, the maximum-entropy value of $|\bm{P}_{\bm{q},\bm{0}}|$ cannot be determined independently of all the other momenta.

To put Eq.~\eqref{eq:tanh} in context, it is helpful to consider the simplification that occurs if we isolate the flavor-mixing part of the system and neglect the entropy associated with the occupation or vacancy of momentum states. The entropy density is then $s_{\bm{q}} = -\textrm{Tr}[\rho_{\bm{q},\bm{0}}\log\rho_{\bm{q},\bm{0}}]$. Holding $U$ fixed, we obtain the following relationship in the maximum-entropy state:
\begin{equation}
    |\bm{P}_{\bm{q},\bm{0}}| = P^0_{\bm{q},\bm{0}} \tanh \beta h_{\bm{q}}. \label{eq:tanh2}
\end{equation}
This result can also be obtained from Eq.~\eqref{eq:tanh} using $|\bm{P}_{\bm{q},\bm{0}}|$, $P^0_{\bm{q},\bm{0}} \ll 1$ for small occupation numbers. A similar equation appears in the mean-field Ising model, relating the magnetization $m$ to the inverse temperature $\beta$ and the effective magnetic field $h$ \cite{callen1980thermodynamics}: 
\begin{equation}
    m = \tanh \beta h.
\end{equation}
This equation, like Eqs.~\eqref{eq:tanh} and \eqref{eq:tanh2}, must be solved self-consistently because $h$ depends on $m$ through the mean-field contribution to $h$ from the spins themselves.

The flavor-temperature hypothesis is analogous to the spin-temperature hypothesis, in which spins weakly coupled to their environment equilibrate via interactions with each other \cite{abragam1958, philippot1964, redfield1969}. The spin-temperature hypothesis has been experimentally validated \cite{purcell1951, abragam1957, abragam1978principles}. Experiments specifically designed to demonstrate its limitations (\textit{e.g.}, in the form of a Loschmidt echo) confirm that the thermalization is fundamentally reversible \cite{rhim1970}, as likewise proposed for neutrinos \cite{johns2023thermodynamics}. Whether a single flavor temperature does or does not emerge is a deep and important fact about neutrino flavor dynamics.

An argument has been made against flavor thermalization on the basis that the kinetic energy of neutrinos is weakly coupled to the refractive sector \cite{fiorillo2025collective}. However, the neutrino kinetic energy (\textit{i.e.}, the energetic contribution from $|\bm{q}|$) is fixed at each momentum because $P^0_{\bm{q}}$ and $\bar{P}^0_{\bm{q}}$ are held constant during the entropy maximization. According to the flavor-temperature hypothesis, thermalization occurs among the flavor polarizations $\langle \bm{P}_{\bm{q}} \rangle$, not between the kinetic-energy and refractive sectors.

Let us stress that neither the adiabatic formulation of miscidynamics presented in Refs.~\cite{johns2023thermodynamics, johns2024subgrid} nor the nonadiabatic extension developed here necessitates flavor thermalization. Miscidynamics evolves the coarse-grained flavor polarizations $|\langle \bm{P}_{\bm{q}} \rangle|$ and makes no assumption about them being related to one another through Eq.~\eqref{eq:tanh} or through the quantum-coherent Fermi--Dirac distribution of Ref.~\cite{johns2023thermodynamics}. On the contrary, flavor thermalization is obstructed by the dynamical emergence of the local oscillation invariants $| \bm{P}_{\bm{q},\bm{0}}|$ and $\epsilon_{\bm{q}}$. In miscidynamics neutrinos are generally expected to be in nonthermal mixing equilibria.

\subsection{Flavor-wave equilibrium}

Consider an isolated region with neutrinos at a stable ($\Omega^{\textrm{I}}_{\bm{n},i} = 0$ for all $\bm{n}$, $i$) mixing equilibrium. Suppose that the system is not subject to any driving, including from collisions, and has $\bm{c}_{\bm{q}} = 0$ for all $\bm{q}$. Under these conditions, the coarse-grained polarization vectors $\langle \bm{P}_{\bm{q}} \rangle$ are constant over any time scale and the state of mixing equilibrium is maintained indefinitely. The flavor-wave amplitudes $|\Phi_{\bm{n},i}|$ may still be evolving, however, due to the nonlinear term $\mathcal{N}^{\textrm{I}}_{\bm{n},i}$ in Eq.~\eqref{eq:ampeom}. In the wave-kinetics approximation, this ongoing subgrid evolution is attributed to three-wave scattering (\textit{i.e.}, triad interactions, with a total of three waves counting ingoing and outgoing).

We give the name \textit{flavor-wave equilibrium} to the end state of this evolution. It is a distinct concept from mixing equilibrium, referring to the flavor-wave spectrum as opposed to the orientations of the coarse-grained polarization vectors. Exact flavor-wave equilibrium requires
\begin{equation}
    \sum_{\bm{m}} \bm{H}_{\bm{q},\bm{n}-\bm{m}} \times \bm{P}_{\bm{q},\bm{m}} = 0
\end{equation}
for all $\bm{q}$ and $\bm{n}$. In flavor-wave kinetics, the equilibrium condition is
\begin{equation}
    \mathcal{N}^{\textrm{I}}_{\bm{n},i} = 0
\end{equation}
for all $\bm{n}$ and $i$. In reality, fluctuations may be significant, and there may be ongoing flavor-wave phase development at fixed amplitude \cite{liu2026dynamical}.

It is conceivable that flavor-wave equilibrium is a maximum-entropy distribution, with flavor waves being driven to this state by three-wave scattering (wave collisions) in the same way that Boltzmann integrals model thermalization via particle collisions. This idea has precedents in other systems exhibiting wave turbulence \cite{zakharov1992kolmogorov, diamond2010modern}. The exponential scaling with $|\bm{k}|$ that is observed in numerical calculations of flavor instabilities \cite{bhattacharyya2020late, bhattacharyya2020fast, wu2021collective, richers2021neutrino, richers2022code, urquilla2024chaos, padilla2025flavor} may be consistent with the maximum-entropy hypothesis. Although these issues are worthy of further investigation, we do not here undertake the extension of neutrino quantum thermodynamics to encompass flavor waves. See Ref.~\cite{fiorillo2025dispersion} for progress in this direction.

A simpler form of nonadiabatic miscidynamics could be formulated based on the assumption of local flavor-wave equilibrium in addition to local mixing equilibrium, obviating the need for full kinetic evolution of the flavor waves. This, too, we leave for future work.

\section{Relation to other work\label{sec:relation}}

We added this section after the first version of the paper appeared, to clarify the connection between it and concurrent work. In hindsight, our flavor-wave transport theory generalizes the quasilinear theory articulated by Fiorillo and Raffelt \cite{fiorillo2024fast, fiorillo2025collective, fiorillo2026quasi}. We will show how our theory reduces to theirs in the specialized case they consider. Much of the present paper has been concerned with the issues that arise in the generalization.

For this section only, to make the comparison easier, we switch to using the notation of their Ref.~\cite{fiorillo2026quasi}. Consider the collisionless fast system with axial symmetry in a periodic box. The (2,1) element of density matrix $\rho_v$ is $\psi_v/2$, with $v$ the cosine of the propagation angle. Through oscillations $\psi_v$ couples to $D_v = (\rho_v)_{11} - (\rho_v)_{22}$. The QKE with spatially homogeneous $D_v$ directly implies Eq.~(7) of Ref.~\cite{fiorillo2026quasi}:
\begin{equation}
    i \partial_t \psi_{v, K} = (kv - D_0) \psi_{v,K} + ( \psi_{0, K} - v \psi_{1,K} ) D_v,\label{eq:psivkEoM}
\end{equation}
where $k \equiv K + D_1$, $D_n \equiv \int dv \, v^n D_v$, and $\psi_n$ is defined similarly. We take Eq.~\eqref{eq:psivkEoM} as the starting point for reproducing their results and demonstrating how our approximation scheme salvages the temporal coarse-graining in less restrictive scenarios.

Suppose that $\psi_{v,K}$ has overlap with at most $M$ discrete collective modes. Expand [compare Eq.~\eqref{eq:Psintexpansion}]
\begin{equation}
    \psi_{v, K} (t) = \sum_{i = 1}^M \mathcal{A}_K^{(i)}(t) e^{-i \vartheta_K^{(i)} (t)} Q_{v,K}^{(i)},\label{eq:psivkexpansion}
\end{equation}
where $Q_{v,K}^{(i)}$ is the $i$th eigenmode and $\mathcal{A}_K^{(i)}$ is the amplitude of this eigenmode after extracting the dynamical phase factor with
\begin{equation}
    \vartheta_K^{(i)}(t) \equiv \int^t dt' \Omega_K^{(i)} (t').
\end{equation}
We will see that the lower boundary of the integral is irrelevant for our purposes. As an instantaneous eigenmode of Eq.~\eqref{eq:psivkEoM}, $Q_{v,K}^{(i)}$ satisfies
\begin{equation}
    \Omega_K^{(i)} Q_{v,K}^{(i)} = (kv - D_0) Q_{v,K}^{(i)} + ( Q_{0, K}^{(i)} - v Q_{1, K}^{(i)} ) D_v\label{eq:OmegaQvk}
\end{equation}
where $\Omega_K^{(i)}$ is its eigenvalue, just as in earlier sections, and
\begin{equation}
    Q_{n,K}^{(i)} \equiv \int_{-1}^{+1} dv \, v^n Q_{v,K}^{(i)}.
\end{equation}
After rearranging terms and defining $\omega_K^{(i)} \equiv \Omega_K^{(i)} + D_0$, we obtain
\begin{equation}
    Q_{v, K}^{(i)} = \frac{Q_{0, K}^{(i)} - v Q_{1, K}^{(i)}}{\omega_K^{(i)} - kv} D_v.
\end{equation}
This relation holds for each eigenmode individually and implies
\begin{equation}
    \psi_{v, K} = \sum_i \frac{\psi_{0, K}^{(i)} - v \psi_{1, K}^{(i)}}{\omega_K^{(i)} - kv} D_v,\label{eq:psivKto0and1}
\end{equation}
using $\psi_{v,K} = \sum_i \psi_{v,K}^{(i)}$ and similar for the angular moments. If $\psi_{v, K}$ has overlap with only a single eigenmode, this reduces to Eq.~(8) of Ref.~\cite{fiorillo2026quasi}. Taking the zeroth angular moment of Eq.~\eqref{eq:OmegaQvk} leads to 
\begin{equation}
    \Omega_K^{(i)} Q_{0, K}^{(i)} = K Q_{1, K}^{(i)}
\end{equation}
and therefore
\begin{equation}
    \psi_{1, K}^{(i)} = \frac{\Omega_K^{(i)}}{K} \psi_{0, K}^{(i)},\label{eq:psi1Krelation}
\end{equation}
which can be used to eliminate $\psi_{1,K}^{(i)}$ in favor of $\psi_{0,K}^{(i)}$. Note that we have only appealed to the eigenmode structure implied by the equation of motion and have not yet made any approximations to the dynamics.

The QKE also directly implies [see their Eq.~(9)]
\begin{equation}
    \partial_t D_v = \sum_K \textrm{Im} \left[ ( \psi_{0, K}^* - v \psi_{1, K}^*) \psi_{v,K}\right].\label{eq:DvEOM}
\end{equation}
Using Eqs.~\eqref{eq:psivKto0and1}, \eqref{eq:psi1Krelation}, and \eqref{eq:DvEOM}, we obtain [compare with our Eq.~\eqref{eq:miscvisc} with flavor-wave viscosity $\gamma_{\bm{q}}$]
\begin{equation}
    \partial_t D_v = -\Gamma_v D_v
\end{equation}
as in Ref.~\cite{fiorillo2026quasi}, but now with the more general flavor-wave viscosity
\begin{widetext}
\begin{equation}
    \Gamma_v \equiv - \sum_K \sum_{i,j} \textrm{Im} \left[ \frac{\left( 1 - v \Omega_K^{(i)*} / K \right) \left( 1 - v \Omega_K^{(j)} / K \right)}{\omega_K^{(j)} - kv} \psi_{0,K}^{(i)*} \psi_{0,K}^{(j)} \right].
\end{equation}
\end{widetext}
In the single-mode case, with $\omega_K = \omega_K^{\textrm{R}} + i \gamma_K$, we obtain
\begin{equation}
    \Gamma_v = \sum_K \frac{\gamma_K}{(\omega_K^{\textrm{R}} - kv)^2 + \gamma_K^2} \left| 1 - \frac{v\Omega_K}{K} \right|^2 \left| \psi_{0,K} \right|^2,
\end{equation}
in agreement with their Eq.~(11). However, we can already see that an issue arises in the multi-mode case that is not encountered under the single-mode approximation: $\Gamma_v$ depends not merely on the scalar $\left| \psi_{0,K} \right|^2$ but on the matrix with elements
\begin{equation}
    \psi_{0,K}^{(i)*} \psi_{0,K}^{(j)} = \mathcal{A}_{K}^{(i)*} \mathcal{A}_{K}^{(j)} e^{i (\vartheta_K^{(i)*} - \vartheta_K^{(j)})} Q_{0,K}^{(i)*} Q_{0,K}^{(j)}.\label{eq:Gammavelements}
\end{equation}
The phase of the exponential factor,
\begin{equation}
    \vartheta_K^{(i)*} - \vartheta_K^{(j)} = \int^t dt' \, ( \Omega_K^{(i)*} - \Omega_K^{(j)}),
\end{equation}
evolves rapidly because a generic flavor-wave energy splitting is on the neutrino oscillation scale. Therefore the fact that Eq.~(11) of Ref.~\cite{fiorillo2026quasi} has no variation at the oscillation scale hinges on the single-mode approximation.

The secular approximation coarse-grains out the rapid phase variation in the more general multi-mode case by discarding all contributions to $\Gamma_v$ except those with $\Omega_K^{(i),\textrm{R}} - \Omega_K^{(j),\textrm{R}} = 0$ [see Eq.~\eqref{eq:rescond0}]. Assuming there are no persistent real-part degeneracies, the secular approximation leads to
\begin{equation}
    \Gamma_v = \sum_K \sum_i \frac{\gamma_K^{(i)}}{(\omega_K^{(i),\textrm{R}} - kv)^2 + (\gamma_K^{(i)})^2} \left| 1 - \frac{v\Omega_K^{(i)}}{K} \right|^2 \left| \psi_{0,K}^{(i)} \right|^2.\label{eq:Gammavnodeg}
\end{equation}
Our theory addresses, among other things, the rapid phase variation of the flavor-wave viscosity $\Gamma_v$ in the multi-mode generalization of Fiorillo and Raffelt's analysis.

Rapid phase variation is not the only issue that arises upon generalizing to multiple modes. Transitions between energy levels need to be considered, even under the quasilinear approximation. To see this, we will consider the flavor-wave equation of motion. First let us introduce the left eigenmodes $\hat{Q}_{v,K}^{(i)}$ satisfying the biorthogonality relation
\begin{equation}
    \langle \hat{Q}_K^{(i)} | Q_K^{(j)} \rangle = \delta_{i,j}
\end{equation} 
with inner product
\begin{equation}
    \langle X | Y \rangle \equiv \int_{-1}^{+1} dv \, X_v^* Y_v
\end{equation}
for two vectors $X_v$ and $Y_v$. Projecting $\partial_t \psi_{v,K}$ onto a single eigenmode and expanding $\psi_{v,K}$ using Eq.~\eqref{eq:psivkexpansion}, we have
\begin{widetext}
\begin{equation}
    \int_{-1}^{+1}dv \, \hat{Q}_{v,K}^{(i)*} \left[ \partial_t \psi_{v,K} \right] = \partial_t \mathcal{A}_K^{(i)} e^{-i \vartheta_K^{(i)}} - i \Omega_K^{(i)} \mathcal{A}_K^{(i)} e^{-i \vartheta_K^{(i)}} + \sum_j \mathcal{A}_K^{(j)} e^{-i \vartheta_K^{(j)}} \langle \hat{Q}_K^{(i)} | \partial_t Q_K^{(j)} \rangle.\label{eq:mathcalA1} 
\end{equation}
\end{widetext}
At the same time, Eqs.~\eqref{eq:psivkEoM} and \eqref{eq:OmegaQvk} imply
\begin{equation}
    \partial_t \psi_{v,K} = - i \sum_j \Omega_K^{(j)} \psi_{v,K}^{(j)}\label{eq:mathcalA1p5}
\end{equation}
and therefore
\begin{equation}
    \int_{-1}^{+1}dv \, \hat{Q}_{v,K}^{(i)*} \left[ \partial_t \psi_{v,K} \right] = -i \Omega_K^{(i)} \mathcal{A}_K^{(i)} e^{-i \vartheta_K^{(i)}}.\label{eq:mathcalA2}
\end{equation}
Equating Eqs.~\eqref{eq:mathcalA1} and \eqref{eq:mathcalA2} results in
\begin{equation}
    \partial_t \mathcal{A}_K^{(i)} = - \sum_j \mathcal{A}_K^{(j)} e^{i ( \vartheta_K^{(i)} - \vartheta_K^{(j)})} \langle \hat{Q}_K^{(i)} | \partial_t Q_K^{(j)} \rangle.\label{eq:partialtmathcalA}
\end{equation}
In the case of a static background, the eigenmodes are time-independent and therefore Eq.~\eqref{eq:partialtmathcalA} gives $\partial_t \mathcal{A} = 0$. The evolution of $\psi$ in this case is entirely captured by the phase factor $e^{-i \vartheta}$. In general, however, time-dependence in the background drives transitions $\mathcal{A}_K^{(i)} \leftrightarrow \mathcal{A}_K^{(j)}$. The quasistatic approximation neglects these transitions, setting $\langle \hat{Q}_K^{(i)} | \partial_t Q_K^{(j)} \rangle = 0$, without imposing the more stringent single-mode approximation used in Ref.~\cite{fiorillo2026quasi}. (Alternatively, we could again appeal to the secular approximation to argue for the suppression of nonquasistatic transitions via phase averaging. We find it useful to distinguish the quasistatic and secular approximations because it may be desirable in some circumstances to apply the first without the second.) Note that the diagonal factors $\langle \hat{Q}_K^{(i)} | \partial_t Q_K^{(i)} \rangle$ can be set to zero using the convention of parallel transport \cite{kost2026neutrino}.

In this section we have assumed, following Fiorillo and Raffelt, that $D_v$ is spatially homogeneous. Nonlinear wave--wave interactions appear when this assumption is relaxed. The right-hand side of Eq.~\eqref{eq:psivkEoM} must be modified to retain the terms
\begin{align}
    \sum_{K' \neq 0} \bigg[& ( \psi_{0, K-K'} - v \psi_{1, K-K'} ) D_{v,K'} \notag\\
    &\hspace{.1 in} - ( D_{0, K'} - v D_{1, K'}) \psi_{v, K - K'} \bigg].\label{eq:FRNLterms}
\end{align}
These neglected terms correspond to our nonlinear terms $N_{\bm{n}}$ [Eq.~\eqref{eq:Nni}]. To calculate them, we need to keep track of $D_{v,K\neq 0}$, whose equation of motion is
\begin{align}
    &\partial_tD_{v,K} = \notag\\
    &\hspace{.2 in} -i K v D_{v,K} + \frac{i}{2} \sum_{K'} \bigg[ (\psi_{0,K'} - v \psi_{1,K'}) \psi_{v, K' - K}^* \notag\\
    &\hspace{0.85 in} - (\psi^*_{0,-K'} - v \psi^*_{1,-K'}) \psi_{v, K - K'} \bigg],\label{eq:DvKNLEoM}
\end{align}
with the sum going over all values of $K'$. According to the traditional linearization in small flavor coherence, the terms in the sum should all be dropped because they are quadratic in $\psi$. In the linearization scheme introduced earlier in this paper, we drop only those terms with $K' \neq 0$ and $K-K' \neq 0$ simultaneously, treating inhomogeneities---but not necessarily coherences---as small. Following this scheme, we can then obtain linearized eigenmodes that generally mix $D$ and $\psi$. Expanding Eq.~\eqref{eq:FRNLterms} in eigenmodes, we again encounter rapid phase variation, as in Eq.~\eqref{eq:Gammavelements}. The secular approximation applies to these nonlinear wave--wave coupling terms as well, as laid out in Sec.~\ref{sec:fwkinetics}.

It is evident from Eq.~\eqref{eq:DvKNLEoM} that $D_{v,K}$ can grow from nonlinear $\psi$--$\psi$ interactions or from collective instabilities. However, $D_{v,K}$ is never unstable if we neglect background values of $\psi_{v,0}$, as has traditionally been done in the literature on linear stability analysis. This can easily be seen from Eq.~\eqref{eq:DvKNLEoM}. After linearizing in $\psi_{v,K \neq 0}$ and dropping $\psi_{v,0}$ terms, we have
\begin{equation}
    \partial_tD_{v,K} = -i K v D_{v,K},
\end{equation}
which only has oscillating solutions. The notion of mixing equilibrium generalizes the idea of slowly evolving background polarizations beyond the trivial configuration where all density matrices $\rho_{v,0}$ are diagonal in the flavor basis. We have defined flavor waves as the inhomogeneous fluctuations on any such background. As we say, collective modes in this more general setting may not be excitations purely in $\psi$, nor purely in $D$, but may in fact mix the two. For this reason we have made no a priori distinction between $\psi$ and $D$, instead packaging them together in our wave functions $\Psi$ [Eq.~\eqref{eq:Psin}].

We arrived at flavor-wave transport theory by considering the backreaction of subgrid fluctuations on slowly evolving local equilibria. In retrospect, the key idea of flavor-wave transport---using flavor-wave eigenanalysis to approximate the evolution of subgrid fluctuations---was first present in Ref.~\cite{fiorillo2024fast}. Our treatment generalizes this theory and shows how it can be used within miscidynamics (which adds the additional approximation of local mixing equilibrium to simplify the $K=0$ evolution). Imposing the single-mode approximation on our framework replicates the quasilinear theory of Fiorillo and Raffelt. Our development of flavor-wave transport theory is an extension in the following key ways.

First, flavor waves $\Psi_{\bm{n}}$ include fluctuations in all polarization components, as opposed to isolating transverse components $\psi_{v, K}$. They evolve on general background ($\bm{n} = 0$) polarization fields rather than backgrounds polarized strictly along the flavor axis, $\langle \bm{P}_{\bm{q}}\rangle \propto \bm{z}$. Flavor coherence is not assumed to be small. The spatially averaged density matrices $\rho_{\bm{q},0}$ are not assumed to be diagonal in the flavor basis.

Second, multiple flavor-wave modes are permitted per wave number $\bm{n}$. The quasistatic and secular approximations are aimed at the complications that arise when there are multiple modes. We have more recently elaborated on the quasistatic approximation and parallel transport in Ref.~\cite{kost2026neutrino}, where the presence of multiple modes was numerically found to be of paramount importance. 

Third, the secular approximation enables the incorporation of nonlinear interactions in a fashion that preserves temporal coarse-graining. We emphasize that the flavor-wave kinetics developed here is distinct from the flavomon kinetics of Ref.~\cite{fiorillo2025collective}. The latter kinetic theory describes resonant interactions involving quantized flavor waves, with \textit{resonant} referring to a $k \cdot v = 0$ condition. Our kinetic theory makes no such restriction. The interactions isolated by the secular approximation are resonant in the sense that classical flavor waves develop phase at the same rate, as in the condition $\Omega_K^{(i),\textrm{R}} - \Omega_K^{(j),\textrm{R}} = 0$. For the purposes of this work, we did not find it useful to quantize the flavor waves.

We use geometric optics to handle the propagation of flavor waves through large-scale inhomogeneous environments. Fiorillo and Raffelt have adopted this approach as well \cite{fiorillo2025neutrino, fiorillo2026flavomon}, arriving at a result in Ref.~\cite{fiorillo2026flavomon} that agrees with our Eq.~\eqref{eq:fwDiagonalLiouville}.

\section{Guide to numerical implementation\label{sec:guide}}

\subsection{Periodic boxes}

In this section we repeat the main results in a streamlined presentation for readers seeking to implement miscidynamics and/or flavor-wave transport numerically. We will first consider the case of flavor evolution in a periodic box, then address the modifications needed for an astrophysical environment with large-scale inhomogeneity.

To evolve the box-averaged density matrices $\langle \rho_{\bm{q}} \rangle$, the miscidynamic equations are solved [Eqs.~\eqref{eq:miscinv}]:
\begin{gather}
    \partial_t \langle P^{0}_{\bm{q}} \rangle = C^{0}_{\bm{q}}, \notag \\
    \partial_t | \langle \bm{P}_{\bm{q}}\rangle |^2 = 2 \langle\bm{P}_{\bm{q}}\rangle \cdot \left( \bm{c}_{\bm{q}} + \bm{C}_{\bm{q}} \right).\label{eq:miscguide}
\end{gather}
The terms $C_{\bm{q}}^0$ and $\bm{C}_{\bm{q}}$ come from decomposing the collision term in the original QKE [Eq.~\eqref{eq:qke}] in the basis of Pauli matrices:
\begin{equation}
    \langle i C_{\bm{q}} \rangle \cong iC_{\bm{q}} [ \langle \rho \rangle ] = \frac{ C_{\bm{q}}^0 + \bm{C}_{\bm{q}} \cdot \bm{\sigma}}{2}.
\end{equation}
The subgrid term $\bm{c}_{\bm{q}}$ is [Eq.~\eqref{eq:cqdefn}]
\begin{equation}
    \bm{c}_{\bm{q}} = \sum_{\bm{n}\neq\bm{0}} \bm{H}_{\bm{q},-\bm{n}} \times \bm{P}_{\bm{q},\bm{n}}.
\end{equation}
At each step, the unit vectors $\langle \bm{P}_{\bm{q}} \rangle / | \langle \bm{P}_{\bm{q}} \rangle|$ are obtained by solving the self-consistency conditions [Eq.~\eqref{eq:mixeqHP}]
\begin{equation}
    \langle \bm{H}_{\bm{q}} \rangle \times \langle \bm{P}_{\bm{q}} \rangle = 0\label{eq:selfconsguide}
\end{equation}
using the vector magnitudes $| \langle \bm{P}_{\bm{q}} \rangle|$ from the miscidynamic equations and enforcing constant-in-time alignment parameters [Eq.~\eqref{eq:epsmixeq}]
\begin{equation}
    \epsilon_{\bm{q}} \equiv \frac{\langle \bm{P}_{\bm{q}} \rangle \cdot \langle \bm{H}_{\bm{q}} \rangle}{|\langle \bm{P}_{\bm{q}} \rangle | | \langle \bm{H}_{\bm{q}} \rangle|} = \pm 1.
\end{equation}
The density matrices can then be reconstructed from the Pauli expansion [Eq.~\eqref{eq:pauli}], or the entire calculation can be carried out in terms of $\langle P_{\bm{q}}^0 \rangle$ and $\langle\bm{P}_{\bm{q}}\rangle$.

This part of the theory---the evolution of $\langle \rho_{\bm{q}} \rangle$---was detailed in the papers introducing miscidynamics \cite{johns2023thermodynamics, johns2024subgrid}. It was then numerically implemented in Ref.~\cite{kost2025once}, where adiabatic ($\bm{c}_{\bm{q}} = 0$) miscidynamics was shown to successfully approximate the coarse-grained evolution in the once-in-a-lifetime encounter model. Sections \ref{sec:mixeq} and \ref{sec:misc} provide additional exposition, but this part of the theory is unchanged from those prior works.

The main new contribution of this paper is a temporally coarse-grained way of treating $\bm{c}_{\bm{q}}$, which captures the effect of subgrid polarization fluctuations $\bm{P}_{\bm{q},\bm{n} \neq 0}$ on $\langle \bm{P}_{\bm{q}}\rangle$. The proposal is flavor-wave transport, wherein flavor waves are propagated using properties determined by the local mean background.

We will describe the implementation for a standard periodic-box calculation with $N$ discrete momenta $\bm{q}_i$. To start, we define the wave vector $\Psi_{\bm{n}}$, packaging together all $\bm{P}_{\bm{q}_i,\bm{n}}$ at a given $\bm{n}$ [Eq.~\eqref{eq:Psin}]. Linearizing the equation of motion in small $\bm{P}_{\bm{q}_i,\bm{n}}$, we deduce the Hamiltonian $\mathcal{H}_{\bm{n}}$ governing the short-time evolution of $\Psi_{\bm{n}}$ [Eq.~\eqref{eq:HnFW}]. The eigensystem of $i\partial_t \Psi_{\bm{n}} = \mathcal{H}_{\bm{n}} \Psi_{\bm{n}}$ is solved at each time step for all $\bm{n}$, and the eigenmode amplitudes $| \Phi_{\bm{n},i} |$ are advanced via [Eq.~\eqref{eq:ampeom}]
\begin{equation}
    \frac{d|\Phi_{\bm{n},i}|}{dt} = 
    \Omega^{\textrm{I}}_{\bm{n},i}|\Phi_{\bm{n},i}| + \mathcal{N}^{\textrm{I}}_{\bm{n},i},\label{eq:Phiguide}
\end{equation}
where $\Omega_{\bm{n},i}$ is the corresponding eigenvalue and superscript I denotes the imaginary part.

The explicit form of the nonlinear term is [Eqs.~\eqref{eq:mathcalNdefn} and \eqref{eq:Nniexpanded}]
\begin{widetext}
\begin{equation}
    \mathcal{N}_{\bm{n},i} = i\sum_{\bm{m} \notin \lbrace \bm{0}, \bm{n}\rbrace} \sum_{j,k,l} |\Phi_{\bm{n}-\bm{m},j}| |\Phi_{\bm{m},l}| e^{i(\varphi_{\bm{n}-\bm{m},j}+\varphi_{\bm{m},l}-\varphi_{\bm{n},i})} [S_{\bm{n}}^{-1}]_{ik} \psi_{\bm{n}-\bm{m},j}^T V_k \psi_{\bm{m},l}.\label{eq:mathcalNexplicit}
\end{equation}
\end{widetext}
The matrix $S_{\bm{n}}$ diagonalizes $\mathcal{H}_{\bm{n}}$ [Eq.~\eqref{eq:Sndiagonalizes}] and the vectors $\psi_{\bm{n},i}$ are the eigenmodes [Eq.~\eqref{eq:Psintexpansion}] as well as the columns of $S_{\bm{n}}$. All of these are obtained when solving the eigensystem. The matrices $V_k$ are defined in Eqs.~\eqref{eq:Vmatx} through \eqref{eq:Vmatz}. The challenge posed by Eq.~\eqref{eq:mathcalNexplicit} is that it contains rapidly varying phases requiring small time steps to resolve. Our proposal for eliminating this rapid motion is to apply the secular approximation, which is implemented numerically by dropping each term in Eq.~\eqref{eq:mathcalNexplicit} not satisfying the resonance condition [Eq.~\eqref{eq:rescond}]
\begin{equation}
    \Omega^{\textrm{R}}_{\bm{n}-\bm{m},j} + \Omega^{\textrm{R}}_{\bm{m},l} - \Omega^{\textrm{R}}_{\bm{n},i} = 0.\label{eq:rescondguide}
\end{equation}
This condition can of course be checked solely using information from the eigenanalysis.

After applying the secular approximation, relative-phase information is left over in $\mathcal{N}$. We retain terms with phases that are stationary over times much less than $t_{\textrm{astro}}$. However, to calculate $\mathcal{N}$, we must know the values of these stationary phases. To handle this we propose a phase-resolved approach, with the phases advanced using [Eq.~\eqref{eq:phaseeom}]
\begin{equation}
    \frac{d\varphi_{\bm{n},i}}{dt}
    = -\Omega^{\textrm{R}}_{\bm{n},i}
    -\frac{\mathcal{N}^{\textrm{R}}_{\bm{n},i}}
    {|\Phi_{\bm{n},i}|}
    \label{eq:varphiguide}
\end{equation}
applying the secular approximation to the nonlinear terms just as in Eq.~\eqref{eq:Phiguide} for the amplitude evolution. The phases then evolve continuously as the background changes, and interactions that subsequently become resonant inherit their relative phases from the preceding evolution. Although the $\Omega$ term in Eq.~\eqref{eq:varphiguide} causes rapid evolution of $\varphi$, it is easily captured numerically because $\Omega$ itself is not rapidly changing. Compare this to absorbing trivial phase evolution in the interaction picture. An alternative would be to adopt a phase-averaged approach by applying a random-phase approximation and evolving only the amplitudes $| \Phi |$. While this avoids tracking the phases, it requires an additional statistical assumption beyond the secular approximation.

The secular approximation allows us to propagate $\Phi_{\bm{n},i}$ using coarse-grained time steps as there is no longer any nontrivial evolution on the scale of $\Delta\Omega \sim \Omega^{\textrm{R}}$. However, flavor waves feed back into the $\langle \rho_{\bm{q}_i} \rangle$ evolution via
\begin{widetext}
\begin{equation}
    c^x_{\bm{q}_i} = \sum_{\bm{n}\neq\bm{0}}\sum_{j,l} |\Phi_{-\bm{n},j}| | \Phi_{\bm{n},l}| e^{i(\varphi_{-\bm{n},j} + \varphi_{\bm{n},l})} \psi_{-\bm{n},j}^T V_{3(i-1)+1} \psi_{\bm{n},l}.\label{eq:cqxexplicit}
\end{equation}
\end{widetext}
The equations for the other vector components $c^y_{\bm{q}_i}$ and $c^z_{\bm{q}_i}$ are identical except with $V_{3(i-1)+1} \rightarrow V_{3(i-1)+2}$ and $V_{3(i-1)+1} \rightarrow V_{3(i-1)+3}$, respectively. Again we apply the secular approximation to eliminate rapidly varying phases, here dropping each term in Eq.~\eqref{eq:cqxexplicit} not satisfying [Eq.~\eqref{eq:rescond0}]
\begin{equation}
    \Omega^{\textrm{R}}_{-\bm{n},j} + \Omega^{\textrm{R}}_{\bm{n},l} = 0.\label{eq:rescond0guide}
\end{equation}

Thus we have a closed system of equations---Eqs.~\eqref{eq:miscguide}, \eqref{eq:Phiguide}, and \eqref{eq:varphiguide} supplemented by the self-consistency condition Eq.~\eqref{eq:selfconsguide}---that approximates neutrino quantum kinetics in a periodic box without resolving all of the small-scale evolution.

Equation~\eqref{eq:Phiguide} adopts the quasistatic approximation, neglecting terms due to finite rates of change of the background. In careful numerical tests, we have found that this approximation is successful away from spectral degeneracies, where special treatment is required (as a last resort, a return to the fine-grained QKEs is possible for those wave numbers $\bm{n}$ afflicted by degeneracies) \cite{kost2026neutrino}. We have not yet performed numerical tests of the secular approximation. Doing so will require comparing the evolution with Eqs.~\eqref{eq:mathcalNexplicit} and \eqref{eq:cqxexplicit} computed exactly against the evolution with all terms dropped that do not satisfy Eqs.~\eqref{eq:rescondguide} and \eqref{eq:rescond0guide}.

\subsection{Global settings}

For flavor waves on slowly varying backgrounds, we advance a formulation in terms of flavor-wave Wigner functions (``density matrices'') $R$ (or $F$ in the propagation basis) rather than the wave vectors $\Psi_{\bm{n}}$. Similar logic and techniques are used to make the transition from quantum field theory to neutrino quantum kinetics \cite{sigl1993general, stirner2018liouville}.

In this phase-space formulation, $F$ is evolved using the quasistatic, linearized ($\mathcal{C}_F = 0$) transport equation [Eq.~\eqref{eq:fwLiouvilleGeometric}] 
\begin{align}
    &(\partial_t + \bm{v}^{\textrm{g}}_{ij}\cdot \partial_{\bm{r}} + \bm{f}_{ij} \cdot\partial_{\bm{k}} )F_{ij} \notag\\
    &\hspace{.6 in}= -i (\Omega_i - \Omega_j^*) F_{ij} - (\mathscr{G}_{ij} + \mathscr{G}_{ji}^*)F_{ij},\label{eq:fwLiouvilleGuide}
\end{align}
with $\Omega$ denoting an eigenvalue of the local flavor-wave Hamiltonian $\mathcal{H}(t,\bm{r},\bm{k})$ [Eq.~\eqref{eq:mathcalHPhaseSpace}], $\bm{v}^{\textrm{g}}_{ij}$ and $\bm{f}_{ij}$ being group velocities and forces [Eqs.~\eqref{eq:groupvelforcedefn} and \eqref{eq:groupvelforceavg}], and $\mathscr{G}$ representing geometric effects associated with the variation of eigenmodes along trajectories in phase space [Eq.~\eqref{eq:mathscrG}].

The flavor waves feed back on the spatially averaged polarization vectors $\langle \bm{P}_{\bm{q}} \rangle$ through the miscidynamic equations [Eqs.~\eqref{eq:miscinv}]
\begin{gather}
    (\partial_t + \bm{\hat{q}}\cdot\partial_{\bm{r}}) \langle P^{0}_{\bm{q}} \rangle = C^{0}_{\bm{q}}, \notag \\
    (\partial_t + \bm{\hat{q}}\cdot\partial_{\bm{r}}) | \langle \bm{P}_{\bm{q}}\rangle |^2 = 2 \langle\bm{P}_{\bm{q}}\rangle \cdot \left( \bm{c}_{\bm{q}} + \bm{C}_{\bm{q}} \right)\label{eq:miscGlobalGuide}
\end{gather}
with [Eqs.~\eqref{eq:fwCgamma} and \eqref{eq:fwgammaqFjl}]
\begin{equation}
    c^a_{\bm{q}_i} = \int \frac{d^3\bm{k}}{(2\pi)^3} \textrm{Tr} \left[ \langle S^{\dagger} V_{3(i-1)+a} S F \rangle \right].\label{eq:cqGlobalGuide}
\end{equation}
The secular approximation is used to average over the rapid evolution in $\bm{c}_{\bm{q}}$. This is implemented by dropping all terms in Eq.~\eqref{eq:cqGlobalGuide} not satisfying the resonance condition [Eq.~\eqref{eq:rescondF0}]
\begin{equation}
    \Omega^{\textrm{R}}_j = \Omega^{\textrm{R}}_l.
\end{equation}
The usual self-consistency conditions [Eq.~\eqref{eq:selfconsguide}] are imposed to ensure local mixing equilibrium.

To test global flavor-wave transport against the QKEs, the initial flavor-wave density matrix $R(t,\bm{r},\bm{k})$ can be evolved as described above after constructing it at $t = t_0$ from the fluctuations $\delta\bm{P}_{\bm{q}} (t_0,\bm{r})$ used as the initial conditions for the QKE calculation. We leave numerical tests to future work.

\section{Summary\label{sec:summ}}

In this study we have extended the theory of miscidynamics by relaxing the adiabatic approximation of Refs.~\cite{johns2023thermodynamics, johns2024subgrid}. These earlier papers proposed a new approach to the problem of neutrino oscillations in extreme astrophysical environments based on a self-consistent coarse-graining of the dynamics. (Self-consistency, as it relates to scale separation and coarse-graining, is discussed in detail in Ref.~\cite{johns2024subgrid}.) The dynamical variables in the adiabatic theory are the coarse-grained density matrices $\langle \rho_{\bm{q}} \rangle(t,\bm{r})$, whose evolution is driven by the astrophysics, \textit{i.e.}, by collisional processes and by propagation in a medium that varies in time and space on the scale of $t_{\textrm{astro}}$. In the nonadiabatic theory, the dynamical variables also include the flavor waves $\Psi_{\bm{n}}(t)$ or their Wigner functions $R(t,\bm{r},\bm{k})$, which interact with $\langle \rho_{\bm{q}} \rangle$ and give rise to the flavor viscosity $\gamma_{\bm{q}}$.

The evolution of $\langle \rho_{\bm{q}} \rangle$, which in general is at the scale of $t_{\textrm{osc}}$, is lifted up to $t_{\textrm{astro}}$ in miscidynamics. This happens for two reasons. First, $\langle \rho_{\bm{q}} \rangle$ tracks local mixing equilibrium, so that by definition the term $\langle \bm{H}_{\bm{q}} \rangle \times \langle \bm{P}_{\bm{q}} \rangle$ causes no rapid variation in $\langle \bm{P}_{\bm{q}} \rangle$. Second, $\gamma_{\bm{q}}$ is small because it arises in response to the astrophysical driving. Thus the terms in the coarse-grained QKE that potentially induce dynamics in $\langle \rho_{\bm{q}} \rangle$ at the $t_{\textrm{osc}}$ scale are actually at the $t_{\textrm{astro}}$ scale, if not weaker. 

If coarse-grained oscillation effects are tied to $t_{\textrm{astro}}$, as we propose, then they are nonvanishing even in the limit of infinite scale separation, $\varepsilon \rightarrow 0$. In particular, $\gamma_{\bm{q}}$ can produce nonadiabaticity in the evolution of $\langle \rho_{\bm{q}} \rangle$ in this limit. It is therefore necessary to distinguish between adiabaticity (no change in $|\langle \bm{P}_{\bm{q}} \rangle|$ due to oscillations) and quasistaticity (no deviation from mixing equilibrium due to a finite rate of driving). These concepts also extend to flavor waves, with the evolution being adiabatic if $|\Phi_{\bm{n},i}|$ undergoes no change due to oscillation terms and quasistatic if level transitions are not induced by the driving. The breakdown of quasistaticity is just one possible cause of nonadiabaticity. It can also be caused by flavor viscosity $\gamma_{\bm{q}}$, flavor-wave growth/damping at rate $\Omega^{\textrm{I}}_{\bm{n},i}$, and three-wave scattering $\mathcal{N}^{\textrm{I}}_{\bm{n},i}$.

Allowing for $\gamma_{\bm{q}} \neq 0$ introduces a closure problem into miscidynamics because $\gamma_{\bm{q}}$ depends on the correlation term $\bm{c}_{\bm{q}}$. Recalling Eq.~\eqref{eq:cqdefn}, we see that $\bm{c}_{\bm{q}}$ contains cross products like
\begin{equation}
    \bm{H}_{\bm{q},-\bm{n}}\times \bm{P}_{\bm{q},\bm{n}}.
\end{equation}
One could imagine complementing the miscidynamic equation for $\langle \rho_{\bm{q}} \rangle$ with an equation of motion for $\gamma_{\bm{q}}$. But then $\partial_t\bm{c}_{\bm{q}}$ appears, which introduces vector triple products like
\begin{equation}
    \bm{H}_{\bm{q},-\bm{n}}\times \left( \bm{H}_{\bm{q},\bm{n}-\bm{m}} \times \bm{P}_{\bm{q},\bm{m}} \right),
\end{equation}
and so on and so forth. 

Reference~\cite{johns2023thermodynamics} observed that nonadiabatic miscidynamics could be formulated by truncating this BBGKY-like hierarchy using the logic of kinetics. The present study follows through on this idea. Under the assumption of weak inhomogeneity, the short-time ($t_{\textrm{osc}}$-scale) evolution of $\Psi$ is governed by a Schr\"{o}dinger equation with non-Hermitian Hamiltonian $\mathcal{H}$. The eigenmodes of $\mathcal{H}$ are the quasiparticles of the system. Imposing the secular approximation leads to kinetic equations coupling the quasiparticles to the neutrinos and to each other. These equations isolate the resonant interactions, which conserve $\bm{k}$ and $\Omega^{\textrm{R}}$. This is similar to how Boltzmann integrals provide a closure by approximating interactions as momentum- and energy-conserving collisions among free particles.

The other question raised by $\gamma_{\bm{q}}$ is how the flavor waves evolve spatially. As an answer to this, we propose that the quasiparticles propagate along rays in a manner resembling the Liouville propagation of neutrinos in quantum kinetics. Properties such as group velocities are determined by the short-time (free-quasiparticle) eigensystem analysis. We have derived this behavior from a gradient expansion of the equation of motion of the flavor-wave Wigner function.

Throughout this work we have adopted the quasistatic approximation. We have done this only for simplicity. Spectral degeneracies may very well be generic. If a gap closes between flavor-wave energy levels, then transitions between these levels become another form of nonadiabaticity that persists even in the limit $\varepsilon \rightarrow 0$. Approximations for non-quasistatic evolution will likely need to be developed.

In this work we have not permitted subgrid inhomogeneity in the astrophysical fluid \cite{abbar2021turbulence, bhattacharyya2025role}, though we expect that it can be accommodated. A related and active debate concerns the question of whether subgrid inhomogeneity needs to be resolved at all, even in the neutrinos \cite{shalgar2022neutrino, cornelius2024, shalgar2024length, liu2025resolution}. In our approach, subgrid inhomogeneities of the neutrino flavor field are contained in the flavor waves $\Psi$. We use the secular approximation to obtain equations of motion without nontrivial small-scale dynamics in time even though the degrees of freedom pertain to small-scale information in space.

In the original presentation of miscidynamics \cite{johns2023thermodynamics}, coarse-graining was performed over both $t$ and $\bm{r}$ in defining $\langle \rho_{\bm{q}}\rangle$. Here we have only coarse-grained over $\bm{r}$. This makes the assumption of local mixing equilibrium more restrictive than it needs to be. In a more general analysis, we could introduce time coarse-graining as well and only insist that $\langle \rho_{\bm{q}} \rangle$ be in local mixing equilibrium on average (referring here to the time average). This generalization leads to nontrivial equilibrium ensembles, which are characterized by different subgrid temporal fluctuations. In this paper we have strictly worked with the trivial ensemble, according to which $\langle \rho_{\bm{q}} \rangle$ has no fine-grained temporal fluctuations. The generalization may be necessary if there are $\bm{n} = \bm{0}$ flavor instabilities driving $\epsilon_{\bm{q}}$ (defined with spatial but not temporal coarse-graining) away from $\pm 1$, as seen in one regime of collisional flavor instability \cite{johns2023collisional}.

The present study has expanded on past explorations of neutrino quantum thermodynamics \cite{johns2023thermodynamics, johns2024subgrid, johns2024ergodicity, johns2023collisional}, proposing that neutrinos generally move through nonthermal mixing equilibria rather than fully flavor-thermalized states. Crucial to this vision is the dynamical emergence of oscillation invariants that are not strictly conserved.

Finally, we have drawn a distinction between mixing equilibria and flavor-wave equilibria, and speculated that the latter may be maximum-entropy states of a suitably defined entropy. Making this idea concrete will require the extension of neutrino quantum thermodynamics to include the entropy associated with the flavor-wave quasiparticles.

\begin{acknowledgments}
We thank Huaiyu Duan, Jiabao Liu, and Hiroki Nagakura for many helpful conversations and Damiano Fiorillo and Georg Raffelt for identifying an error in our implementation of flavor-wave geometric optics in an earlier version of this manuscript. L. J. is supported by a Feynman Fellowship through LANL LDRD project number 20230788PRD1. A. K. is supported by the US DOE NP grant No. DE-SC0017803 at UNM. We used ChatGPT-6 to assist in proofreading the manuscript.
\end{acknowledgments}

\bibliography{refs}
\bibliographystyle{apsrev4-2}

\end{document}